%% file: Waveguide-BlochPBC.tex
\documentclass{pj}
\usepackage{amssymb}
\usepackage{graphics}
\usepackage[dvips]{graphicx}
\usepackage{latexsym,amsmath,amsfonts,amscd,subfigure}
\usepackage{epstopdf}
\usepackage{epsfig}
\usepackage{changebar}
\usepackage{indentfirst}
\usepackage{verbatim}
\usepackage{graphicx}
\usepackage{amsmath,amsthm}
\usepackage{extarrows}

\input latex_defs.tex

\begin{document}
\setcounter{page}{1}
\pjheader{Vol.\ x, y--z, 2018}

\title[Mixed Spectral Element Method for the Waveguide Problem with Bloch Periodic Boundary]
{Mixed Spectral Element Method for the Waveguide Problem with Bloch Periodic Boundary}
 \footnote{\it Received date}
 \footnote{\hskip-0.12in*\, Corresponding
author: Qing Huo Liu (qhliu@duke.edu).}
\footnote{\hskip-0.12in\textsuperscript{1} Institute of Electromagnetics and Acoustics, College of Electronic Science and Technology, Xiamen University, Xiamen 361005, China. \\
\textsuperscript{2} Hubei Subsurface Multi-scale Imaging Key Laboratory, Institute of Geophysics and Geomatics, China University of Geosciences, Wuhan 430074, China.\\
\textsuperscript{3} Shenzhen Research Institute, Xiamen University, Shenzhen 518057, China.\\
\textsuperscript{4} Department of Electrical and Computer Engineering,
           Duke University, Durham, NC, 27708, USA.}

\author{Jie Liu\textsuperscript{1}, Wei Jiang\textsuperscript{2}, Na Liu\textsuperscript{1,3} and Qing Huo Liu\textsuperscript{*,1,4}}

\runningauthor{Jie Liu, Wei Jiang, Na Liu and Qing Huo Liu}

\tocauthor{Institution of Electromagnetics and Acoustics, Xiamen, 361005, P.R. China.  and FistName1~LastName1}

\begin{abstract}
The mixed spectral element method (MSEM) is applied to solve the waveguide problem with Bloch periodic boundary condition (BPBC). Based on the BPBC for the original Helmholtz equation and the periodic boundary condition (PBC) for the equivalent but modified Helmholtz equation, two equivalent mixed variational formulations are applied for the MSEM. Unlike the traditional finite element method and SEM, both these mixed SEM schemes are completely free of spurious modes because of their use of the Gauss' law and the curl-conforming vector basis functions structured by the Gauss-Legendre-Lobatto (GLL) points. A simple implementation method is used to deal with the BPBC and the PBC for the mixed variational formulations so that both schemes can save computational costs over the traditional methods. Several numerical results are also provided to verify that both schemes are free of spurious modes and have high accuracy with the propagation constants.
\end{abstract}

\setlength {\abovedisplayskip} {6pt plus 3.0pt minus 4.0pt}
\setlength {\belowdisplayskip} {6pt plus 3.0pt minus 4.0pt}

\section{Introduction}
\label{section label}
Various types of waveguides are widely used in the fields of microwave and optical technologies, in which the eigen analysis is an important research topic. The waveguide eigenvalue problem is to determine the propagation constants and the corresponding guided modes in a given waveguide configuration. If a guided wave structure is canonical and filled with an isotropic homogeneous medium, the exact eigenpair can be obtained by an analytical method. However, when the guided wave structure is irregular and/or is filled with complex (such as anisotropic and/or inhomogeneous) media, it is difficult to obtain an analytical solution for the waveguide problem. Hence, an effective and highly accurate numerical method for the waveguide problem is necessary. The main numerical methods for the waveguide problems include the finite difference method \cite{Guan1995, Xu2003}, the finite element method (FEM) \cite{Fujisawa2017, Lee1991, Schulz2003, Gentili2016}, the method of moments \cite{Lai2011}, the multidomain pseudospectral method \cite{Song2015}, the multipole method \cite{Mohamed1992}, and the mode matching method \cite{QHLiu1991a,QHLiu1991b,Selleri2001}, and so forth.

For a numerical method, we mostly pursue its accuracy, efficiency and correctness. In this regard, the spectral element method (SEM), which combines the advantages of the high accuracy of the spectral method and the geometric flexibility of the FEM, is becoming more and more popular. It is well known that the Legendre polynomial (LP) and the Chebyshev polynomial (CP) can minimize the Runge phenomenon and the singularity of the solutions near the boundary or at the interface between different media. The SEM has the basis functions constructed by the high-degree orthogonal LPs or CPs, so that it not only achieves spectral accuracy, but also greatly reduces the computational costs compared with the conventional high-order FEM \cite{Lee2006}. In general, at the sampling density of 4 points per wavelength (PPW), the SEM can achieve a numerical error smaller than $0.1\%$ for an appropriate smooth solution \cite{Luo2009}. Therefore, it has been applied in various fields; for instance, photonic/phononic crystals \cite{Luo2009, Luo2009x, Shi2016}, biharmonic equations \cite{Shen1994}, fluid dynamics \cite{Canuto2007}, elastic waves \cite{Komatitsch1998,Liu2017} and acoustic waves \cite{Shi2016x} and the subsurface electromagnetics \cite{Zhou2017}, and so on. Particularly, the spectral element method has been also used for the waveguide problems in \cite{Peverini2011} and \cite{Liu2015}.

In addition, spurious modes in numerical methods are often mixed with physical modes of a waveguide problem. In \cite{Vardapetyan2002}, a novel variational formulation with the N\'{e}d\'{e}lec element and the Gauss' law was developed to eliminate all the spurious modes. The mixed finite element methods (MFEM), which is composed of the curl-conforming basis functions and the Gauss' law, was also employed to suppress the spurious modes \cite{Chen2004,Liu2016}. Within the framework of the FEM and using the formulation of \cite{Vardapetyan2002}, Liu \emph{et al.} \cite{Liu2015} provided the mixed spectral element method (MSEM) to solve the dielectric waveguide problems. The MSEM is completely free of spurious modes and has exponential convergence because the Guss-Lobatto-Legendre (GLL) polynomials are applied to construct the curl-conforming vector edge-based basis functions for the transverse electric field, the scalar continuous nodal-based basis functions for the longitudinal component, and Gauss' law is enforced.

In many research fields, the waveguide problem with the Bloch periodic boundary conditions (named here as the BPBC waveguide problem) plays a crucial role, for instance, the fiber Bragg grating, optical lithography and metasurfaces. For these problems, in order to obtain accurate electromagnetic fields, we need to directly/indirectly solve the BPBC waveguide problem. For example, the electromagnetic fields can be expressed in term of an expansion of waveguide eigenmodes in the numerical mode-matching (NMM) method \cite{QHLiu1990}. However, on our knowledge, few articles focus on this type of waveguide problem. Thus, it is meaningful to calculate the propagation constants and the corresponding guided modes of the BPBC waveguide problem with an efficient and accurate method free of spurious modes.

This paper is devoted to the high-accuracy numerical solutions for the BPBC waveguide problem without any spurious modes. Based on the BPBC and the periodic boundary condition (PBC), there are two equivalent mixed variational formulations for the MSEM. A simple implementation method is developed to deal with the BPBC and PBC for the mixed variational formulations so that both schemes can save computational costs over the traditional methods. Numerical examples indicate that both schemes are completely free of spurious modes and have high accuracy.

The rest of this paper is organized as follows. In Section 2, the governing equations and two variational formulations are introduced. The basis functions and the discrete forms are presented in Section 3. In Section 4, the accuracy and efficiency of both MSEM schemes are demonstrated by several numerical results.

\section{Governing Equations and Variational Formulations}
\label{sec:formulation}
\subsection{Governing Equations}
The waveguide problem is also known as a 2.5-dimensional problem because of its fields are three-dimensional but the medium properties $(\bar{\bar{\epsilon}}_{r}, \bar{\bar{\mu}}_{r})$ are only two-dimensional. Assume that the propagation is along the $+z$-direction and the cross section $\Gamma$ of the waveguide is uniform in the $z$-direction. The phasor expressions for the electric field $\textbf{E}$ and the magnetic field $\textbf{H}$ can be written as:
\begin{equation}\label{eq:1}
\left\{
  \begin{array}{ll}
    \textbf{E}(x,y,z)=\textbf{e}(x,y)e^{-jk_{z}z}\\
    \textbf{H}(x,y,z)=\textbf{h}(x,y)e^{-jk_{z}z},
  \end{array}
\right.
\end{equation}
where $\textbf{e}(x,y)$ and $\textbf{h}(x,y)$ are two-dimensional vector phasors that depend only on the transverse coordinates $(x,y)$, and $k_{z}$ is the propagation constant along the $+z$-direction.

We next write the fields and the operator $\nabla$ in terms of the transverse components and the $z$ components, that is
\begin{equation}\label{eq:2}
\left\{
  \begin{array}{lll}
    \textbf{e}(x,y)=\textbf{e}_{t}+\hat{z}e_{z}\\
    \textbf{h}(x,y)=\textbf{h}_{t}+\hat{z}h_{z}\\
    \nabla=\nabla_{t}+\hat{z}\frac{\partial}{\partial z}=\nabla_{t}-\hat{z}jk_{z}.
  \end{array}
\right.
\end{equation}
Let the medium parameters be the following forms:
\begin{equation}\label{eq:3}
\bar{\bar{\epsilon}}_{r}=
\begin{bmatrix} \bar{\bar{\epsilon}}_{rt} &0\\0&\epsilon_{rz}\end{bmatrix},
\bar{\bar{\mu}}_{r}=
\begin{bmatrix} \bar{\bar{\mu}}_{rt} &0\\ 0&\mu_{rz}\end{bmatrix},
\end{equation}
where $\bar{\bar{\epsilon}}_{rt}$ and $\bar{\bar{\mu}}_{rt}$ are the transversal parts of the relative permittivity and the relative permeability tensors, respectively, $\epsilon_{rz}$ and $\mu_{rz}$ are longitudinal parts, and they are independent of $z$.
Substituting (\ref{eq:1})-(\ref{eq:3}) into the source-free Maxwell's equations, and eliminating the magnetic fields $\textbf{h}_{t}$ and $h_{z}$, we arrive at
\begin{subequations}\label{eq:4}
\begin{align}
\nabla_{t}\times\mu_{rz}^{-1}\nabla_{t}\times\textbf{e}_{t}+jk_{z}\hat{z}\times\bar{\bar{\mu}}_{rt}^{-1}\hat{z}\times(\nabla_{t}e_{z}+jk_{z}\textbf{e}_{t})
-k_{0}^{2}\bar{\bar{\epsilon}}_{rt}\textbf{e}_{t}=0\\
\nabla_{t}\times\bar{\bar{\mu}}_{rt}^{-1}\hat{z}\times(\nabla_{t}e_{z}+jk_{z}\textbf{e}_{t})+k_{0}^{2}\epsilon_{rz}e_{z}\hat{z}=0\\
\nabla_{t}\cdot(\bar{\bar{\epsilon}}_{rt}\textbf{e}_{t})-jk_{z}\epsilon_{rz}e_{z}=0,
\end{align}
\end{subequations}
where $k_{0}$ is the wave number in vacuum.

In order to suppress these spurious modes, we here follow the argument of MSEM \cite{Liu2015} to employ equation (4a) and the divergence condition (4c) as the governing equations.

To facilitate the operability of the subsequent numerical calculations, we first introduce the following rotation matrix:
\begin{equation}\label{eq:5}
\bar{\bar{R}}=
\begin{bmatrix} \cos(\pi/2) &-\sin(\pi/2)\\\sin(\pi/2)&\cos(\pi/2)\end{bmatrix}
=\begin{bmatrix} 0 &-1\\1&0\end{bmatrix},
\end{equation}
which is equivalent to the operator $\hat{z}\times$, with the property $\bar{\bar{R}}^{2}=-\bar{\bar{I}}$, where $\bar{\bar{I}}$ denotes a unit matrix;
and then a new variable $e_{z}^{new}$ is defined by $e_{z}^{new}=jk_{z}e_{z}$. Inserting (\ref{eq:5}) into (4a), the governing equations can be obtained from (4a) and (4c)
\begin{subequations}\label{eq:6}
\begin{align}
\nabla_{t}\times\mu_{rz}^{-1}\nabla_{t}\times\textbf{e}_{t}+\bar{\bar{R}}\bar{\bar{\mu}}_{rt}^{-1}\bar{\bar{R}}\nabla_{t}e_{z}^{new}
-k_{0}^{2}\bar{\bar{\epsilon}}_{rt}\textbf{e}_{t}
=k_{z}^{2}\bar{\bar{R}}\bar{\bar{\mu}}_{rt}^{-1}\bar{\bar{R}}\textbf{e}_{t}\\
\nabla_{t}\cdot(\bar{\bar{\epsilon}}_{rt}\textbf{e}_{t})-\epsilon_{rz}e_{z}^{new}=0.
\end{align}
\end{subequations}
In general, $\bar{\bar{R}}\bar{\bar{\mu}}_{rt}^{-1}\bar{\bar{R}}$ is not equal to $-\bar{\bar{\mu}}_{rt}^{-1}$ unless $\bar{\bar{\mu}}_{r}$ has a special forms (see, eq. (26) in \cite{Jiang2017}). Consequently, (\ref{eq:6}) is a more general form for the waveguide problem than the one of \cite{Liu2015}.

In order to solve (\ref{eq:6}), one usually need to provide suitable boundary conditions; for example, the PEC boundary, the PMC boundary and the radiation boundary condition, etc. However, to treat periodic waveguides such as photonic-crystal waveguides, we here focus on the following Bloch periodic boundary conditions:
\begin{equation}\label{eq:7}
\textbf{e}_{t}(\textbf{r}+\textbf{a})=\textbf{e}_{t}(\textbf{r})e^{-j\textbf{k}_{t}\cdot\textbf{a}},~
e_{z}^{new}(\textbf{r}+\textbf{a})=e_{z}^{new}(\textbf{r})e^{-j\textbf{k}_{t}\cdot\textbf{a}},
\end{equation}
where $\textbf{k}=\textbf{k}_{t}+\hat{z}k_{z}$ is the Bloch wave vector, $\textbf{r}$ and $\textbf{a}$ are the position vector on the boundary $\partial\Gamma$ and the lattice translation vector, respectively.

\subsection{Variational Formulations}

To construct the variational formulations of the waveguide problem (\ref{eq:6}), we take the inner
product of the differential equations (6a) and (6b) with appropriate test functions $\textbf{v}(\textbf{r})$ and $q(\textbf{r})$, respectively, and integrate by parts to obtain
\begin{equation}\label{eq:8}
\begin{split}
k_{0}^{2}c(\textbf{e}_{t},\textbf{v})-a(\textbf{e}_{t},\textbf{v})
+b(\nabla_{t}e_{z}^{new},\textbf{v})
+I_{1}
=k_{z}^{2}b(\textbf{e}_{t},\textbf{v})
\end{split}
\end{equation}
and
\begin{equation}\label{eq:9}
d(\textbf{e}_{t},q)
-I_{2}
+m(e_{z}^{new},q)=0,
\end{equation}
where the boundary terms $I_{1}\equiv\int_{\partial\Gamma}\hat{\textbf{n}}\times\textbf{v}^{*}\cdot\mu_{rz}^{-1}\nabla_{t}\times\textbf{e}_{t}d\ell$, $I_{2}\equiv\int_{\partial\Gamma}q^{*}\hat{\textbf{n}}\cdot\bar{\bar{\epsilon}}_{rt}\textbf{e}_{t}d\ell$ with the unit outward normal $\hat{\textbf{n}}$ at a point $\textbf{x}=(x,y)$ on the edge $\ell\in\partial\Gamma$ and the symbol '$*$' denotes conjugate transpose. The above bilinear forms can be written as:
\begin{eqnarray}\label{eq:10}
a(\textbf{e}_{t},\textbf{v})&=&\int_{\Gamma}(\nabla_{t}\times\textbf{v})^{*}\cdot\mu_{rz}^{-1}\nabla_{t}\times\textbf{e}_{t}d\textbf{x}\nonumber\\
b(\textbf{e}_{t},\textbf{v})&=&-\int_{\Gamma}\textbf{v}^{*}\cdot\bar{\bar{R}}\bar{\bar{\mu}}_{rt}^{-1}\bar{\bar{R}}\textbf{e}_{t}d\textbf{x}\nonumber\\
c(\textbf{e}_{t},\textbf{v})&=&\int_{\Gamma}\textbf{v}^{*}\cdot\bar{\bar{\epsilon}}_{rt}\textbf{e}_{t}d\textbf{x}\\
d(\textbf{e}_{t},q)&=&\int_{\Gamma}(\nabla_{t}q)^{*}\cdot\bar{\bar{\epsilon}}_{rt}\textbf{e}_{t}d\textbf{x}\nonumber\\
m(e_{z}^{new},q)&=&\int_{\Gamma}q^{*}\epsilon_{rz}e_{z}^{new}d\textbf{x}.\nonumber
\end{eqnarray}

\begin{figure}[!t]
  \centering
{
   \includegraphics[width=0.45\columnwidth,draft=false]{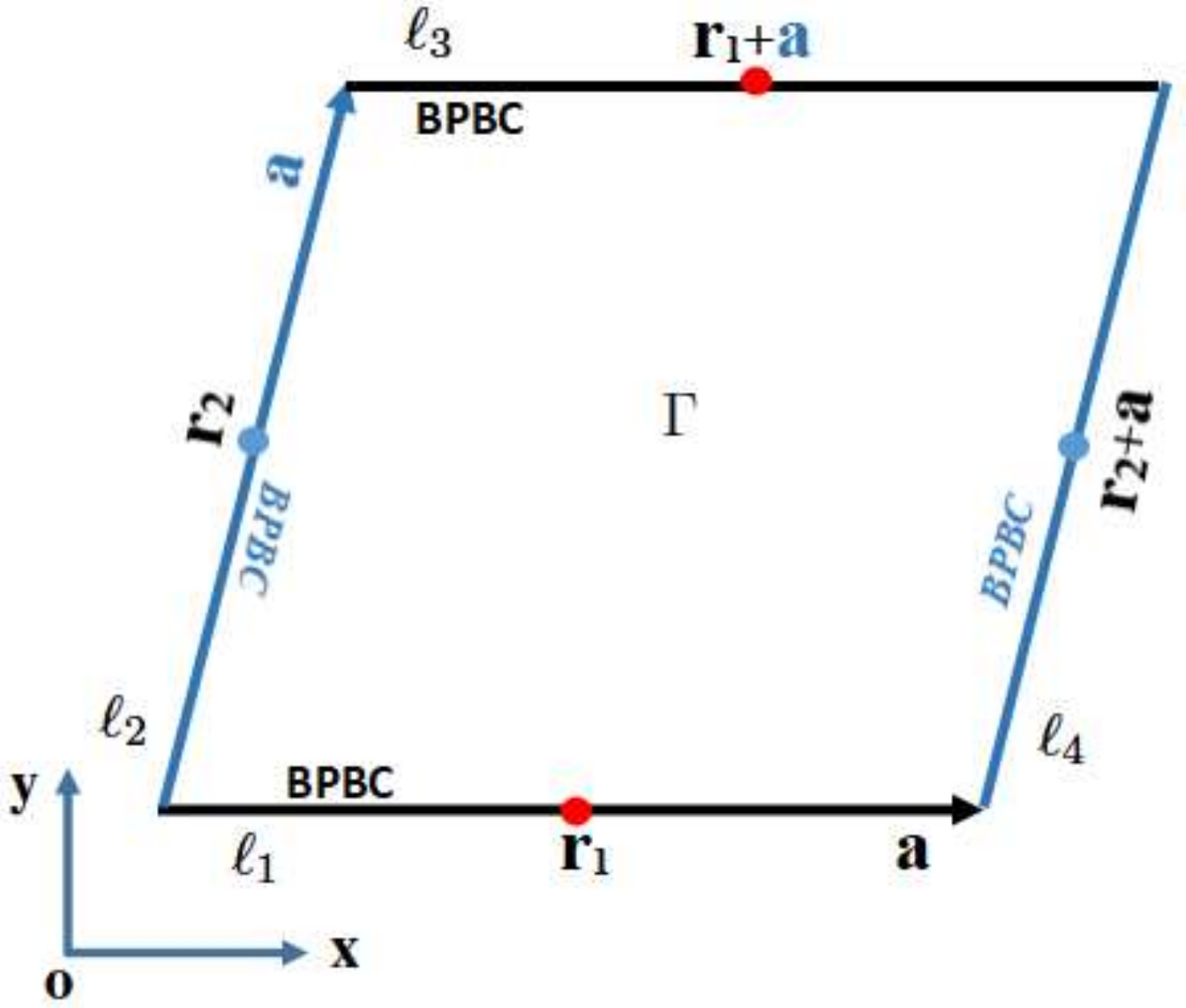}
   }
 \caption{Sketch map for the cross section $\Gamma$ of BPBC waveguide.}
\label{sketch3}
\end{figure}

Let $\{\ell_{k},\ell_{k+2}\}\subset\partial\Gamma$, for $k=1,2$, be a pair of Bloch periodic boundary (see, Fig. \ref{sketch3}). We next consider two boundary integrations:
\begin{equation}\label{eq:11}
\int\limits_{\ell_{k}}\hat{\textbf{n}}_{k}\times\textbf{v}^{*}(\textbf{r})\cdot\mu_{rz}^{-1}\nabla_{t}\times\textbf{e}_{t}(\textbf{r})d\ell
+\int\limits_{\ell_{k+2}}\hat{\textbf{n}}_{k+2}\times\textbf{v}^{*}(\textbf{r}+\textbf{a})\cdot\mu_{rz}^{-1}\nabla_{t}\times\textbf{e}_{t}(\textbf{r}+\textbf{a})d\ell,
\end{equation}
\begin{equation}\label{eq:12}
\begin{split}
\int\limits_{\ell_{k}}q^{*}(\textbf{r})\hat{\textbf{n}}_{k}\cdot\bar{\bar{\epsilon}}_{rt}\textbf{e}_{t}(\textbf{r})d\ell
+\int\limits_{\ell_{k+2}}q^{*}(\textbf{r}+\textbf{a})\hat{\textbf{n}}_{k+2}\cdot\bar{\bar{\epsilon}}_{rt}\textbf{e}_{t}(\textbf{r}+\textbf{a})d\ell.
\end{split}
\end{equation}
We first restrict $q$ and $\textbf{v}$ to the Bloch periodic subspaces, respectively
\begin{equation*}
H_{p}^{B}(\Gamma)
=\{q\in H^{1}(\Gamma):q(\textbf{r+\textbf{a}})=q(\textbf{r})e^{-j\textbf{k}_{t}\cdot\textbf{a}}~\textrm{on} ~\partial\Gamma\},
\end{equation*}
\begin{equation*}
\textbf{H}_{p}^{B}(\textrm{curl};\Gamma)
=\{\textbf{v}\in\textbf{H}(\textrm{curl};\Gamma):\textbf{v}(\textbf{r+\textbf{a}})=\textbf{v}(\textbf{r})e^{-j\textbf{k}_{t}\cdot\textbf{a}}~\textrm{on} ~\partial\Gamma\}
\end{equation*}
consistent with the Bloch periodic conditions (\ref{eq:7}), and then the boundary integrations (\ref{eq:11}) and (\ref{eq:12}) vanish due to the fact $\hat{\textbf{n}}_{k}=-\hat{\textbf{n}}_{k+2}$. Consequently, the boundary terms $I_{1}=I_{2}=0$ in (\ref{eq:8}) and (\ref{eq:9}) when $\{\ell_{k},\ell_{k+2}\}$ traverse all the boundary edges on $\partial\Gamma$.

The mixed variational formulation of (\ref{eq:6}) reads:  Given a $k_{0}$, find $(k_{z},\textbf{e}_{t},e_{z}^{new})\in \mathbb{C}\times\textbf{H}_{p}^{B}(\textrm{curl};\Gamma)\times H_{p}^{B}(\Gamma)$, for all $\textbf{v}\in\textbf{H}_{p}^{B}(\textrm{curl};\Gamma)$, such that
\begin{subequations}\label{eq:13}
\begin{align}
k_{0}^{2}c(\textbf{e}_{t},\textbf{v})-a(\textbf{e}_{t},\textbf{v})+b(\nabla_{t}e_{z}^{new},\textbf{v})=k_{z}^{2}b(\textbf{e}_{t},\textbf{v})\\
d(\textbf{e}_{t},q)+m(e_{z}^{new},q)=0, ~\forall q\in H_{p}^{B}(\Gamma).
\end{align}
\end{subequations}

Alternatively, if the eigenfunctions $\textbf{e}_{t}$ and $e_{z}^{new}$ are written as the plane wave forms
\begin{equation}\label{eq:14}
\textbf{e}_{t}(\textbf{k}_{t},\textbf{r})=\textbf{u}(\textbf{k}_{t},\textbf{r})e^{-j\textbf{k}_{t}\cdot\textbf{r}},
e_{z}^{new}(\textbf{k}_{t},\textbf{r})=w(\textbf{k}_{t},\textbf{r})e^{-j\textbf{k}_{t}\cdot\textbf{r}}
\end{equation}
by the Bloch theorem, the periodic boundary conditions can be obtained from (\ref{eq:7})
\begin{equation}\label{eq:15}
\textbf{u}(\textbf{k}_{t},\textbf{r})=\textbf{u}(\textbf{k}_{t},\textbf{r}+\textbf{a}),~
w(\textbf{k}_{t},\textbf{r})=w(\textbf{k}_{t},\textbf{r}+\textbf{a}),
\end{equation}
which indicates the wavefunctions $\textbf{u}(\textbf{k}_{t},\textbf{r})$ and $w(\textbf{k}_{t},\textbf{r})$ are periodic functions. Substituting (\ref{eq:14}) into (\ref{eq:11}) and (\ref{eq:12}), with the periodicity (\ref{eq:15}) and the identity $\hat{\textbf{n}}_{k}=-\hat{\textbf{n}}_{k+2}$, it is easy to check that
both the boundary integrations are still zero, only the operator $\nabla_{t}$ is replaced by operator $\nabla_{t}-j\textbf{k}_{t}$.

We then achieve a mixed variational formulation which is equivalent to (\ref{eq:13}) as follows: Given a $k_{0}$, find $(k_{z},\textbf{u},w)\in \mathbb{C}\times\textbf{H}_{p}(\textrm{curl};\Gamma)\times H_{p}(\Gamma)$, for all $\textbf{v}\in\textbf{H}_{p}(\textrm{curl};\Gamma)$, such that
\begin{subequations}\label{eq:16}
\begin{align}
k_{0}^{2}\tilde{c}(\textbf{u},\textbf{v})-\tilde{a}(\textbf{u},\textbf{v})+\tilde{b}((\nabla_{t}-j\textbf{k}_{t})w,\textbf{v})=k_{z}^{2}\tilde{b}(\textbf{e}_{t},\textbf{v})\\
\tilde{d}(\textbf{u},q)+\tilde{m}(w,q)=0, ~\forall q\in H_{p}(\Gamma).
\end{align}
\end{subequations}
Compared to (\ref{eq:13}), only the operator $\nabla_{t}-j\textbf{k}_{t}$ comes into the bilinear forms with the tilde instead of the operator $\nabla_{t}$ in (\ref{eq:10}); the periodic subspaces $H_{p}(\Gamma)$ and $\textbf{H}_{p}(\textrm{curl};\Gamma)$ derive from the Bloch periodic subspaces $H_{p}^{B}(\Gamma)$ and $\textbf{H}_{p}^{B}(\textrm{curl};\Gamma)$ in the absence of the Bloch factor $e^{-j\textbf{k}_{t}\cdot\textbf{a}}$, respectively. Equations (\ref{eq:16}) transform the BPBC waveguide problem (\ref{eq:13}) into the PBC waveguide problem. In our numerical experiments, both schemes are used.

\section{Basis Functions and Discretization}

\subsection{Basis Functions}

As the description in \cite{Lee2006,Luo2009,Luo2009x, Shi2016, Shi2016x, Zhou2017} and \cite{Liu2015}, the $\emph{N}$th-order GLL polynomials can be used to interpolate a suitable smooth function with spectral accuracy.
We here use the Lagrange interpolation basis functions associated with the GLL sample points $\xi_{k}\in[-1,1]$ as the $\emph{N}$th-order 1D GLL basis functions, for all $j=0,1,\ldots,N$
\begin{equation}\label{eq:17}
\phi_{j}^{(N)}(\xi)=\prod_{\substack{k=0\\k\neq j}}^{N}\frac{\xi-\xi_{k}}{\xi_{j}-\xi_{k}},~\forall \xi\in[-1,1],
\end{equation}
where the GLL points $\xi_{k}$ are the $(N+1)$ roots of the equation $(1-\xi_{k}^{2})L'_{N}(\xi_{k})=0$, and $L'_{N}(\xi)$ is the derivative of the $\emph{N}$th-order Legendre polynomial $L_{N}(\xi)$.

For our 2.5-D problem, the scalar field $e_{z}^{new}$ is interpolated by the tensor-product nodal basis function
\begin{equation}\label{eq:18}
\tilde{\psi}_{ij}(\xi,\eta)=\phi_{i}^{(N)}(\xi)\phi_{j}^{(N)}(\eta), ~\forall (\xi, \eta)\in[-1,1]^{2},
\end{equation}
and the transversal component $\textbf{e}_{t}$ can be approximated by the curl-conforming vector edge-based basis functions
\begin{equation}\label{eq:19}
\bm{\tilde{\Phi}}_{i,j}^{\xi}=\hat{\xi}\phi_{i}^{(N-1)}(\xi)\phi_{j}^{(N)}(\eta),
\end{equation}
\begin{equation}\label{eq:20}
\bm{\tilde{\Phi}}_{i,j}^{\eta}=\hat{\eta}\phi_{i}^{(N)}(\xi)\phi_{j}^{(N-1)}(\eta).
\end{equation}
In order to easily calculate the elemental matrices by using the above basis functions in the reference element $\hat{\kappa}=[-1,1]^2$, the invertible mappings $x(\xi,\eta)$ and $y(\xi,\eta)$ (see, eq. (18) in \cite{Luo2009}) are first employed to transform the physical element $\kappa$ (which may be curved) into $\hat{\kappa}$, and then the corresponding covariant mappings \cite{Monk2003} are applied to the basis functions $\psi$ and $\bm{\Phi}$ in $\kappa$, that is
\begin{equation}\label{eq:21}
\left\{
  \begin{array}{llll}
  \psi(x,y)=\tilde{\psi}(\xi,\eta)\\
  \bm{\Phi}(x,y)=\textbf{J}^{-1}\bm{\tilde{\Phi}}(\xi,\eta)\\
  \nabla\psi(x,y)=\textbf{J}^{-1}\hat{\nabla}\tilde{\psi}(\xi,\eta)\\
  \nabla\times\bm{\Phi}(x,y)=\frac{1}{|\textbf{J}|}\textbf{J}^{T}\hat{\nabla}\times\bm{\tilde{\Phi}}(\xi,\eta)\\
  \end{array}
\right.
\end{equation}
where the Jacobian matrix $\textbf{J}$ is defined by
\begin{equation*}
\textbf{J}=
\begin{bmatrix} \frac{\partial x}{\partial\xi} &\frac{\partial y}{\partial \xi}& 0\\
                \frac{\partial x}{\partial\eta}&\frac{\partial y}{\partial \eta}&0 \\
                0 & 0 &1
\end{bmatrix}.
\end{equation*}
Define the spectral element spaces $Q^{N,h}$ and $\textbf{V}^{N,h}$ by
\begin{equation*}
Q^{N,h}=\textrm{span}\{\psi_{1},\psi_{2},\ldots,\psi_{M_{n}}\},
\end{equation*}
and
\begin{equation*}
\textbf{V}^{N,h}=\textrm{span}\{\bm{\Phi}_{1},\bm{\Phi}_{2},\ldots,\bm{\Phi}_{M_{e}}\},
\end{equation*}
where the $M_{n}$ is the total nodal degree of freedom (DOF) and $M_{e}$ denotes the total edge-DOF.

\subsection{Discrete Forms}
Without loss of generality, for the BPBC waveguide problem, we follow the standard numerical process of the FEM.
First, let the longitudinal component and the transverse vector components of the electric field within $\kappa$ be expressed as
\begin{equation}\label{eq:22}
(e_{z}^{new})^{N,h}|_{\kappa}=\sum_{r,s=0}^{N_{n}}e_{r,s}\psi_{r,s},~
\textbf{e}_{t}^{N,h}|_{\kappa}=\sum_{l,r,s}^{N_{e}}E_{l,r,s}\bm{\Phi}_{r,s}^{l},
\end{equation}
where $N_{n}=(N+1)^2$ is the local nodal DOF, $N_{e}=2N(N+1)$ is the local edge-DOF, the superscript $l$ denotes the corresponding directions of $\xi$ and $\eta$ in the physical space.
Meanwhile, for any $(e_{z}^{new})^{N,h}\in Q^{N,h}$ and $\textbf{e}_{t}^{N,h}\in \textbf{V}^{N,h}$, $\mathbf{E}$ can be written as
\begin{equation}\label{eq:23}
(e_{z}^{new})^{N,h}=\sum_{i=1}^{M_{n}}f_{i}\psi_{i},~
\textbf{e}_{t}^{N,h}=\sum_{k=1}^{M_{e}}F_{k}\bm{\Phi}_{k},
\end{equation}
where $i=\{r,s\}$ and $k=\{l,r,s\}$ are the compound index of the basis functions.

On the other hand, substituting (\ref{eq:23}) into (\ref{eq:13}) and (\ref{eq:16}), by taking $\textbf{v}=\bm{\Phi}_{k}$ and $q=\psi_{i}$,
we obtain the generalized waveguide eigenvalue problems, respectively
\begin{equation}\label{eq:24}
\begin{bmatrix}
k_{0}^{2}\bar{\bar{C}}-\bar{\bar{A}}& \bar{\bar{B}}_{1}\\
\bar{\bar{D}} & \bar{\bar{M}}
\end{bmatrix}
\begin{bmatrix}
\textbf{F}\\
\textbf{f}
\end{bmatrix}
=(k_{z}^{N,h})^{2}
\begin{bmatrix}
\bar{\bar{B}}_{2}& O\\
O & O
\end{bmatrix}
\begin{bmatrix}
\textbf{F}\\
\textbf{f}
\end{bmatrix},
\end{equation}

\begin{equation}\label{eq:25}
\begin{bmatrix}
k_{0}^{2}\tilde{\bar{C}}-\tilde{\bar{A}}& \tilde{\bar{B}}_{1}\\
\tilde{\bar{D}} & \tilde{\bar{M}}
\end{bmatrix}
\begin{bmatrix}
\textbf{F}\\
\textbf{f}
\end{bmatrix}
=(k_{z}^{N,h})^{2}
\begin{bmatrix}
\tilde{\bar{B}}_{2}& O\\
O & O
\end{bmatrix}
\begin{bmatrix}
\textbf{F}\\
\textbf{f}
\end{bmatrix},
\end{equation}
where $k_{z}^{N,h}$ is the approximate eigenvalue (propagation constant or wave number along the $z$-direction); $\textbf{F}=[F_{1},F_{2},\ldots, F_{M_{e}}]^{T}$ and $\textbf{f}=[f_{1},f_{2},\ldots,f_{M_{n}}]^{T}$ are the eigenvectors of the approximate fields $\textbf{e}_{t}^{N,h}$ and $(e_{z}^{new})^{N,h}$, respectively, which can satisfy the boundary conditions (\ref{eq:7}) or (\ref{eq:15}). The corresponding elemental matrices can be obtained by inserting (\ref{eq:22}) into (\ref{eq:13}) and (\ref{eq:16}) with the mappings (\ref{eq:21}). For example, the elemental matrices $\bar{\bar{A}}^{(\kappa)}$ and $\tilde{\bar{A}}^{(\kappa)}$ consist of the following parts, respectively
\begin{equation*}
\begin{split}
&a(\bm{\tilde{\Phi}}_{i},\bm{\tilde{\Phi}}_{k})_{\hat{\kappa}}=\int_{\hat{\kappa}}\frac{1}{|\textbf{J}|}(\textbf{J}^{T}\hat{\nabla}_{t}\times\bm{\tilde{\Phi}}_{k}^{*})\cdot\mu_{rz}^{-1}\textbf{J}^{T}\hat{\nabla}_{t}\times\bm{\tilde{\Phi}}_{i}d\xi d\eta\\
&\tilde{a}(\bm{\tilde{\Phi}}_{i},\bm{\tilde{\Phi}}_{k})_{\hat{\kappa}}=\int_{\hat{\kappa}}\frac{1}{|\textbf{J}|}(\textbf{J}^{T}(\hat{\nabla}_{t}+j\textbf{k}_{t})\times\bm{\tilde{\Phi}}_{i}^{*})
\cdot\mu_{rz}^{-1}\textbf{J}^{T}(\hat{\nabla}_{t}-j\textbf{k}_{t})\times\bm{\tilde{\Phi}}_{i}d\xi d\eta.
\end{split}
\end{equation*}

\subsection{Imposing Boundary Conditions}

We impose the BPBC (\ref{eq:7}) and the PBC (\ref{eq:15}) on (\ref{eq:24}) and (\ref{eq:25}), respectively. For the PEC boundary, it is well known that we only need to remove the corresponding rows and columns of the matrices in (\ref{eq:24}) according to the numbering which represents the DOF lying on the PEC boundary. Along this way, here we shall do some minor adjustments for the Bloch periodic boundary. As an example, let matrix $\bar{\bar{A}}$ be
\begin{equation*}
\bar{\bar{A}}=
\begin{bmatrix} a_{11} &a_{12}& \cdots & a_{1,M_{e}}\\
                a_{21} &a_{22}& \cdots & a_{2,M_{e}}\\
                \vdots&\vdots&\ddots&\vdots\\
                a_{M_{e},1}& a_{M_{e},2}&\cdots& a_{M_{e},M_{e}}
\end{bmatrix},
\end{equation*}
where $a_{i,k}=a(\bm{\tilde{\Phi}}_{i},\bm{\tilde{\Phi}}_{k})_{\hat{\kappa}}$, for all $i,k=1,2,\ldots M_{e}$. Meanwhile, we assume that the first and $M_{e}$-th DOF lying on the Bloch periodic boundaries (of course, it can be any pair), i.e., $F_{M_{e}}=F_{1}e^{-j\textbf{k}_{t}\cdot\textbf{a}}$. Thus, we can obtain from (\ref{eq:23})
\begin{equation}\label{eq:26}
\textbf{e}_{t}^{N,h}=F_{1}\bm{\Phi}'_{1}+\sum_{k=2}^{M_{e}-1}F_{k}\bm{\Phi}_{k},
\end{equation}
where $\bm{\Phi}'_{1}=\bm{\Phi}_{1}+\bm{\Phi}_{M_{e}}$. Applying the bilinear form $a(\cdot,\cdot)_{\hat{\kappa}}$ and (\ref{eq:26}), we have
\begin{equation*}
\bar{\bar{A}}'=
\begin{bmatrix}
 a(\bm{\Phi}'_{1},\bm{\Phi}'_{1})_{\hat{\kappa}} &a(\bm{\Phi}'_{1},\bm{\Phi}_{2})_{\hat{\kappa}}& \cdots & a(\bm{\Phi}'_{1},\bm{\Phi}_{M_{e}-1})_{\hat{\kappa}}\\
                a(\bm{\Phi}_{2},\bm{\Phi}'_{1})_{\hat{\kappa}} &a_{22}& \cdots & a_{2,M_{e}-1}\\
                \vdots&\vdots&\ddots&\vdots\\
                a(\bm{\Phi}_{M_{e}-1},\bm{\Phi}'_{1})_{\hat{\kappa}}& a_{M_{e}-1,2}&\cdots& a_{M_{e}-1,M_{e}-1}
\end{bmatrix}.
\end{equation*}
By a simple calculation, we find that $\bar{\bar{A}}$ and $\bar{\bar{A}}'$ satisfy the following relation:
\begin{equation}\label{eq:27}
\bar{\bar{A}}\xLongrightarrow[(C_{1}+C_{M_{e}}e^{-j\textbf{k}_{t}\cdot\textbf{a}})/C_{M_{e}}]{(R_{1}+R_{M_{e}}e^{j\textbf{k}_{t}\cdot\textbf{a}})/R_{M_{e}}}\bar{\bar{A }}'
\end{equation}
where $R_{k}$ and $C_{k}$ denote the $k$th row and column of a matrix ($k=1,M_{e}$), respectively; the symbol '$/$' indicates that a row or column of a matrix is deleted. In a similar way, the BPBC (\ref{eq:7}) can be imposed on all the matrices of (\ref{eq:24}). For the PBC (\ref{eq:15}), it can be processed to all the matrices of (\ref{eq:25}), by only taking $\textbf{a}=\textbf{0}$ in (\ref{eq:27}). In short, our both MSEM schemes not only have the favorable inheritance, but also reduce the computational costs by using (\ref{eq:27}) to remove part of the degrees of freedom. Note that we still use (\ref{eq:24}) and (\ref{eq:25}) to express the waveguide eigenvalue problems which have been imposed the BPBC (\ref{eq:7}) and the PBC (\ref{eq:15}), respectively.

\section{Numerical Experiments}

In this section, we will report some examples for solving the BPBC waveguide problems by the MSEM, and validate them with the commercial FEM solver COMSOL. We will compare the CPU time, the number of degrees of freedom and the accuracy for our schemes and COMSOL to show the high accuracy and efficiency of our schemes. We use Matlab R2010b and COMSOL to solve (\ref{eq:24}) and (\ref{eq:25}) on a PC. Note that COMSOL will produce many nonphysical modes, therefore only the physical modes are listed in our tables.

In our numerical experiments, the wave vector $\textbf{k}$ is defined by
\begin{equation}\label{eq:28}
\textbf{k}=k(\hat{x}\sin\theta\cos\phi+\hat{y}\sin\theta\sin\phi+\hat{z}\cos\theta),
\end{equation}
where $k=k_{0}\sqrt{\mu_{r}\epsilon_{r}}$, $(\theta,\phi)$ are the elevation and azimuthal angles of the propagation direction. The CPU time listed is the average time obtained by running our codes and COMSOL three times for the same experiment on a ThinkPadT450 PC. In the tables below, $k_{i,z}^{N,h}$ denotes the $i$-th approximate eigenvalue achieved by the MSEM with $N$ order basis functions. The solutions $k_{i,z}^{10,h}$ are used as the reference values. Note that our DOF includes the nodal DOF and the edge-DOF, i.e., $M_{n}+M_{e}$.

\subsection{Fiber Bragg Grating}

The fiber Bragg grating (FBG) is a fiber with the periodically patterned refractive index in the core. An alternative approach is that generating a periodic embossing on the surface of the waveguide such that the refractive index is periodically patterned.
By the Bloch theory, the electric and magnetic fields are assumed to consist of an infinite number of space harmonics when an electromagnetic wave propagates in a periodic structure of a dielectric waveguide; for instance, the tangential fields can be written as:
\begin{equation}\label{eq:29}
\textbf{E}_{t}=\sum_{\alpha=1}^{\infty}\textbf{e}_{\alpha,t}e^{-\gamma_{\alpha}z}f_{\alpha}(z)
\end{equation}
where $\gamma_{\alpha}=jk_{\alpha,z}$ is the $\alpha$-th propagation constant of the BPBC waveguide problem, $\textbf{e}_{\alpha,t}$ is the corresponding guided mode and $f_{\alpha}(z)$ is the amplitude function. It is easy to see that for given $f_{\alpha,s}$, the field $\textbf{E}_{t}$ depends on the wave number $k_{\alpha,z}$ and the guided modes $\textbf{e}_{\alpha,t}$.

\begin{figure}[h]
  \centering
{
   \includegraphics[width=0.4
   \columnwidth,draft=false]{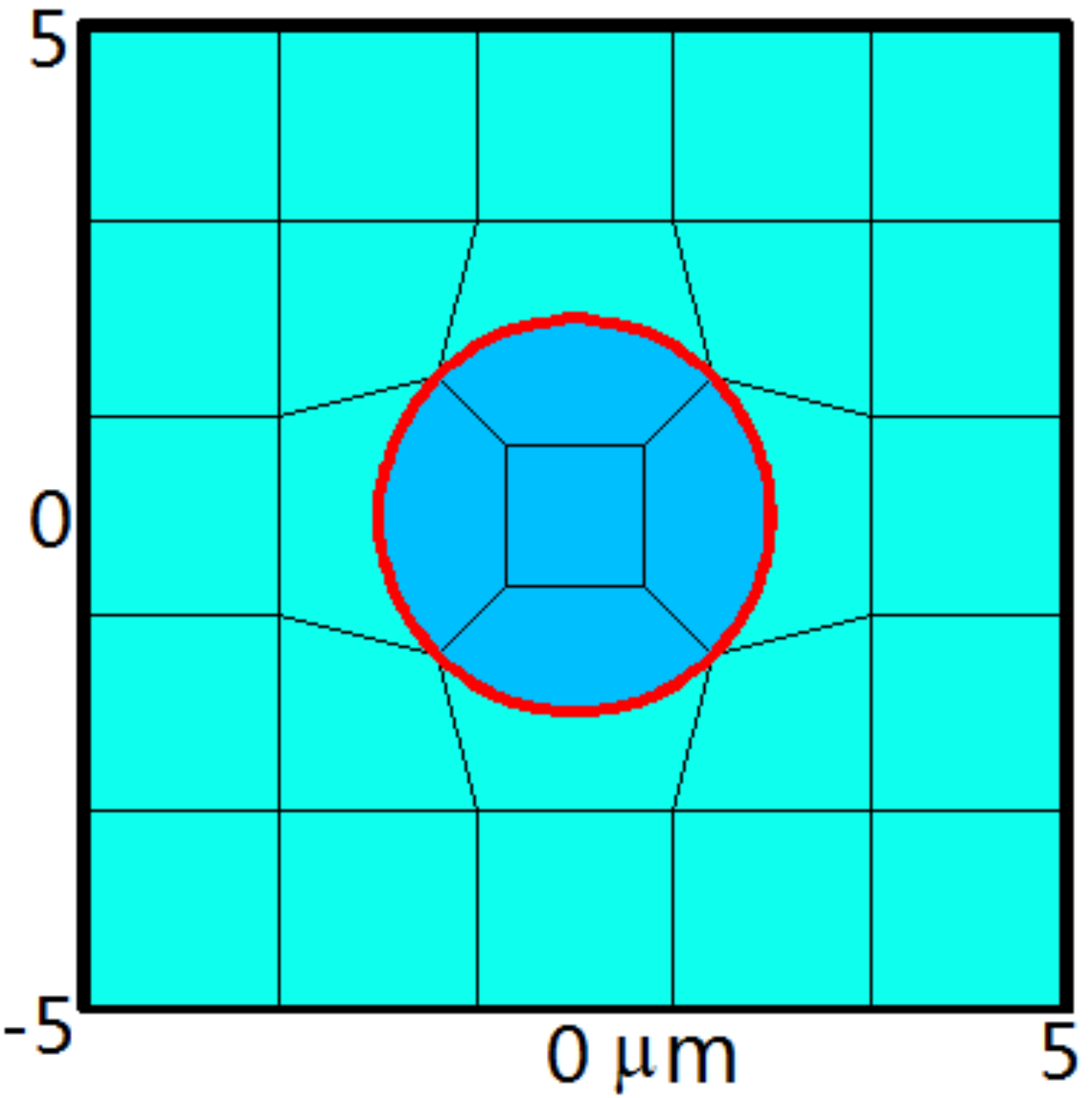}
   }
 \caption{A unit cell of FBG and mesh for the MSEM.}
\label{sketch1}
\end{figure}

A unit cell and its mesh are shown in Fig. \ref{sketch1}; the lattice constant is 10 $\mu m$, the circular cladding is silica with a refractive index $n_{\textrm{SiO}_{2}}=1.46$ and the radius $2~\mu m$ in air. The elevation and azimuthal angles $(\theta,\phi)=(\pi/6,\pi/3)$ for the wave vector $\textbf{k}$. The operating wavelength is $5~\mu m$ and $k=k_{0}$.  In this example, the 6th-order MSEM $(N=6)$ and the 2nd-order FEM are applied to solve (\ref{eq:24}) and (\ref{eq:25}) on the quadrangular mesh (see, Fig. \ref{sketch1}) and a triangular mesh in COMSOL, respectively.

In order to obtain accurate solutions, from Table \ref{example11}, we can see that COMSOL requires more 34.96 times DOFs and 7.758 times CPU time than the MSEM. Also, we can see that our schemes (\ref{eq:24}) and (\ref{eq:25}) have nearly the same accuracy which indicates that (\ref{eq:24}) is equivalent to (\ref{eq:25}). In fact, using the algorithm (\ref{eq:24}), we can arrive at the reference value $k_{i,z}^{10,h}$ by only taking $N=6$.

In practice, the core of the FBG may be filled with a lossy medium. To verify that our schemes are accurate and efficient for the inhomogeneous anisotropic lossy medium, we consider the following reciprocal medium for the cladding
\begin{equation}\label{eq:30}
\bar{\bar{\epsilon}}_{r}=
\begin{bmatrix}
1 & 1-0.5j & 0\\
1-0.5j & 2-j &0\\
0  & 0& 3
\end{bmatrix}.
\end{equation}
The elevation and azimuthal angles of the propagation direction are changed to $(\theta,\phi)=(\pi/4,\pi/4)$.

\begin{figure}[h]
  \centering
  {\subfigure[]{
    \label{value1}
   \includegraphics[width=0.35\columnwidth,draft=false]{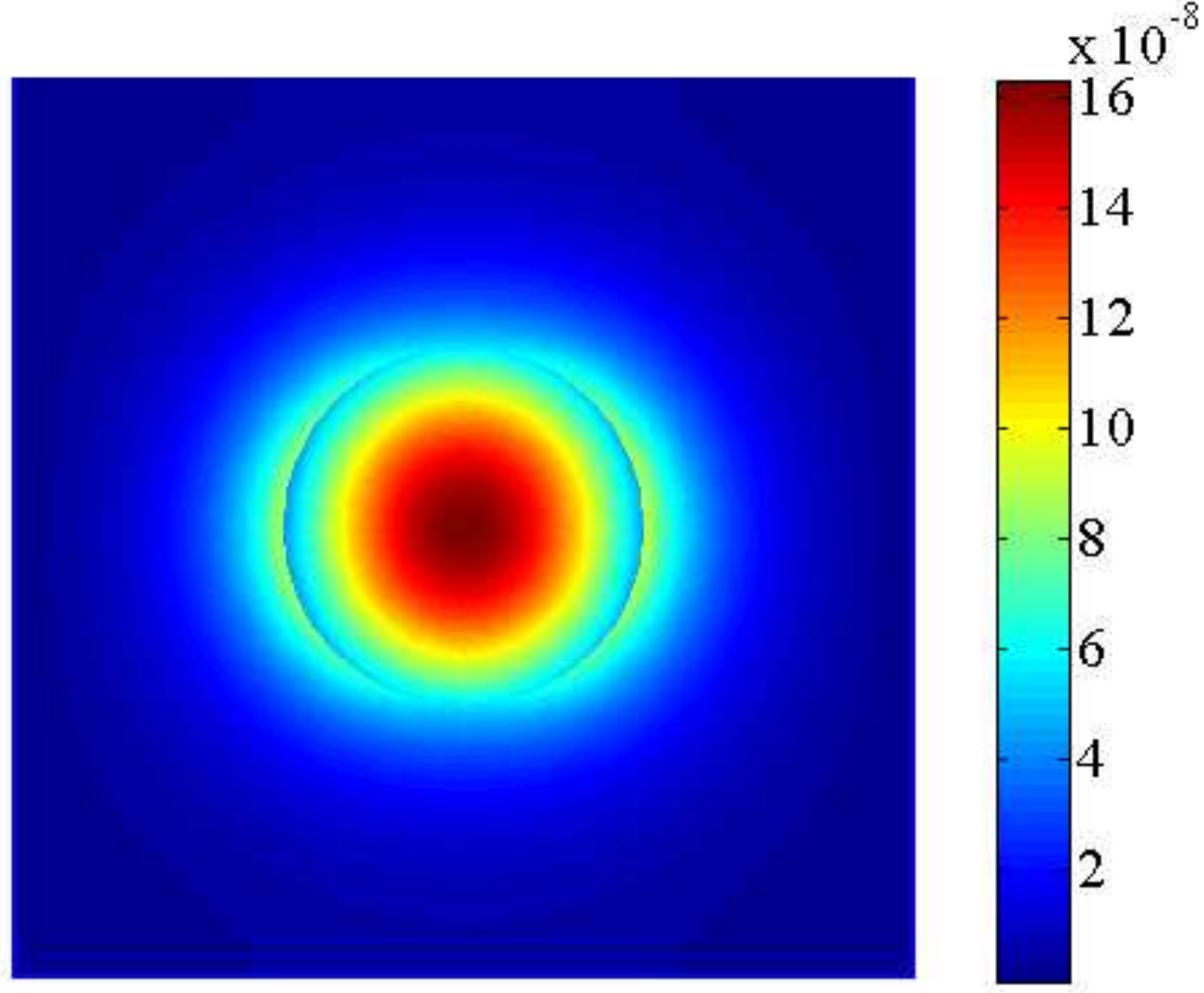}}
     \subfigure[]{
    \label{value2}
   \includegraphics[width=0.35\columnwidth,draft=false]{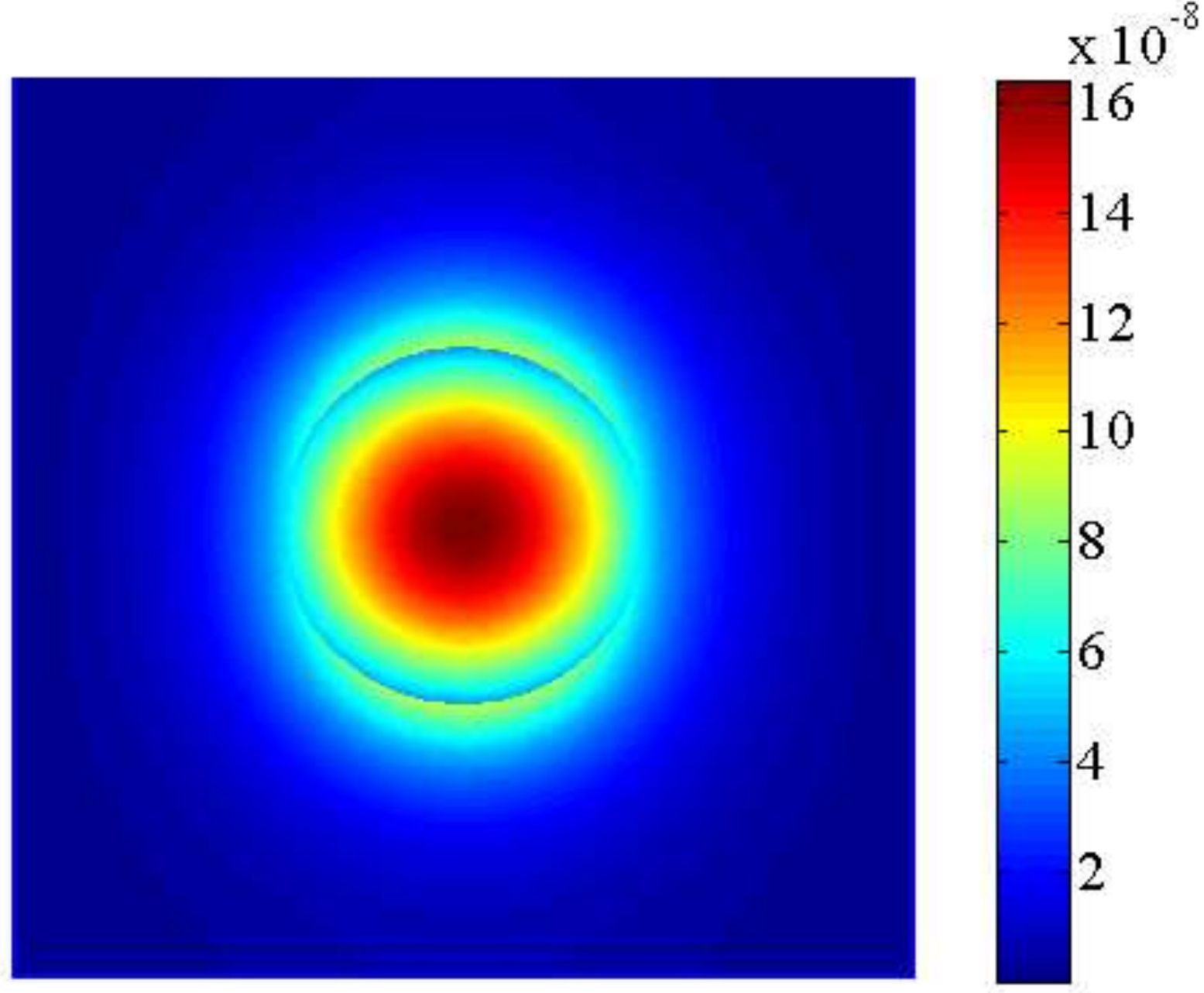}}
   \subfigure[]{
    \label{value3}
   \includegraphics[width=0.35\columnwidth,draft=false]{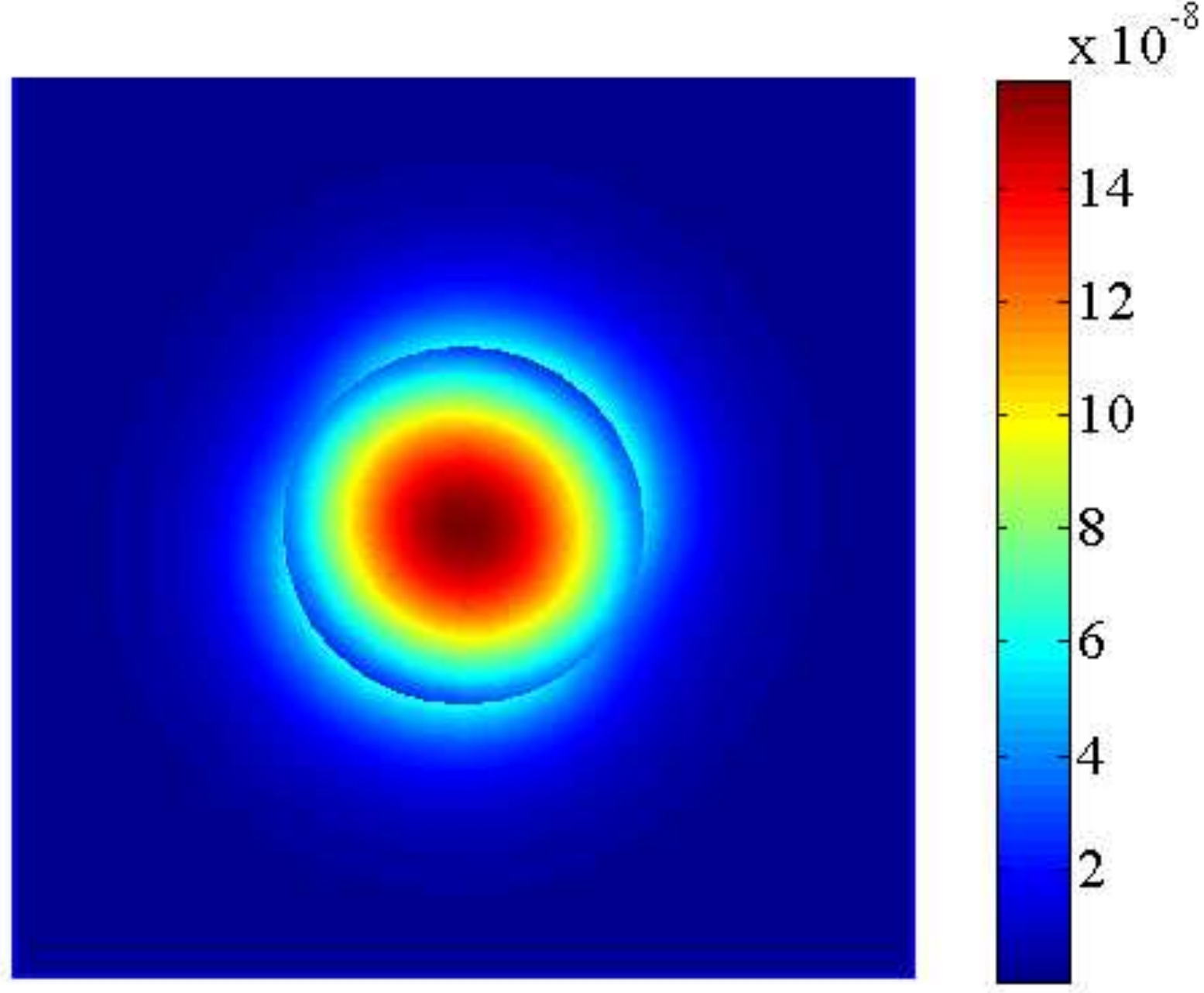}}
     \subfigure[]{
    \label{value4}
   \includegraphics[width=0.35\columnwidth,draft=false]{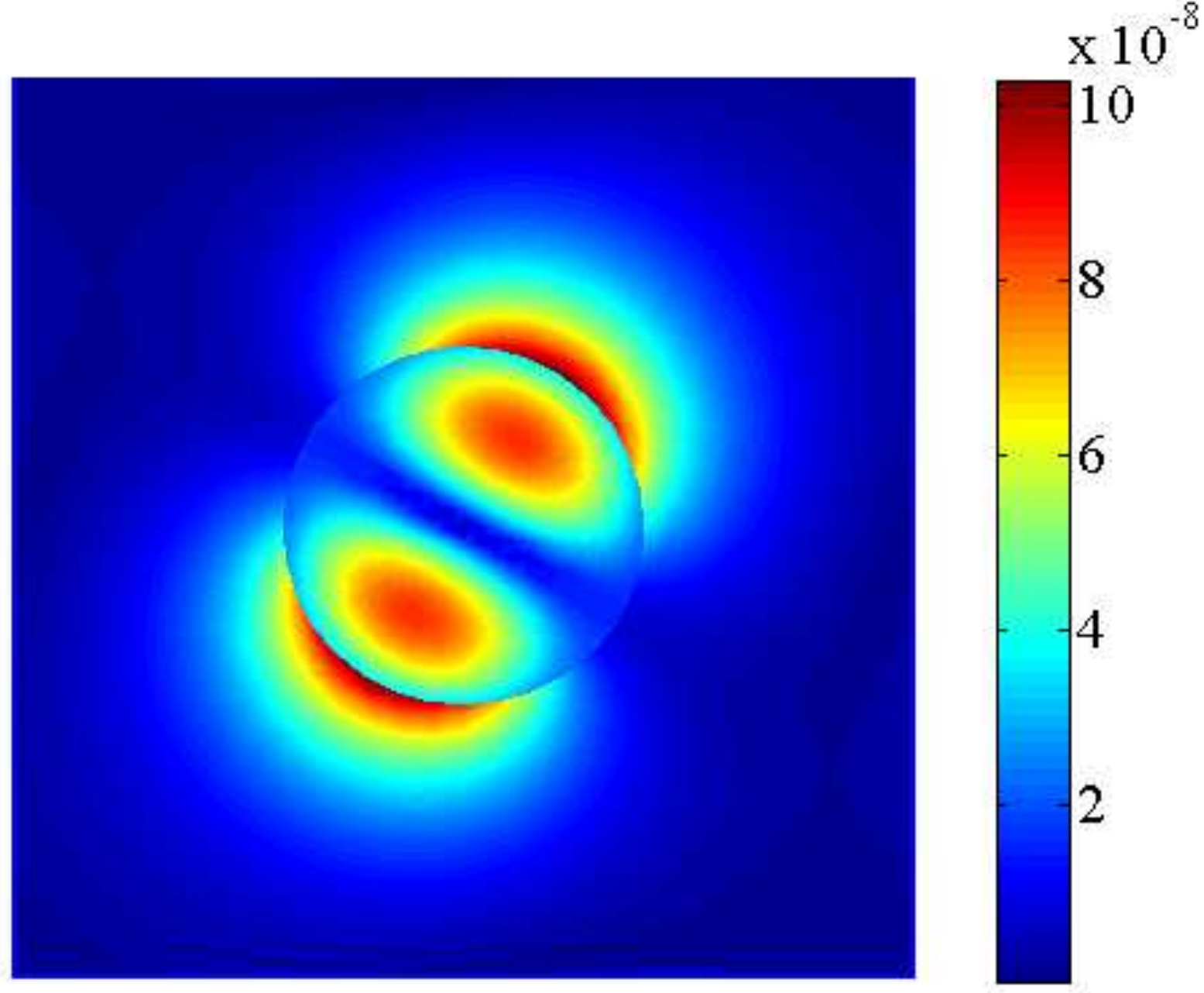}}
   }
 \caption{Magnitude distributions of the field $\textbf{e}_{t}^{N,h}$ corresponding to the first two eigenvalues. (a) and (b) for the  $\textrm{SiO}_{2}$ core. (c) and (d) for the core with the lossy medium (\ref{eq:30}).}
\label{field1}
\end{figure}

In Table \ref{example12}, we present the numerical results in the presence of the reciprocal medium. First, it is easy to observe that the similar accuracy solutions can be also obtained by our schemes and by COMSOL with the same DOFs and mesh as the above example; however, COMSOL spends 9.017 times longer CPU time than MSEM; our algorithm (\ref{eq:24}) can arrive at  $k_{i,z}^{10,h}$ when $N=6$. Furthermore, for the lossy medium, we find that the spurious modes, which are generally plural or zero, are confusing when screening for the physical modes in COMSOL, because the physical modes are also plural; yet our schemes (\ref{eq:24}) and (\ref{eq:25}) are completely free of all the spurious modes. Finally, in Fig. \ref{field1}, we show the magnitude distributions of the electric field $\textbf{e}_{t}^{N,h}$ corresponding to the first two propagation modes in the case of the core with $\textrm{SiO}_{2}$ and with lossy medium (\ref{eq:30}), respectively.

\begin{table}[h]
\caption{Results on the BPBC waveguide with the $\textrm{SiO}_{2}$ core obtained by the MSEM and COMSOL}
\vskip0.2in
\centering
\begin{tabular}{ccccc}
\hline
$k_{i,z}^{N,h}$ &Eqn (\ref{eq:24})&Eqn (\ref{eq:25})&\textrm{COMSOL}& $k_{i,z}^{10,h}$\\
\hline
1&  1582575.03 &1582574.98 &1582575.13 &1582575.04\\

2&  1582571.26 &1582571.16 &1582571.35 &1582571.26\\

3&  1310557.88 &1310557.53 &1310558.06 &1310557.93\\

4&  1287648.24 &1287647.92 &1287648.30 &1287648.22\\
\hline
Time(s)&5.797 & 6.359    &49.333     &27.469\\

DOF    &3132    &3132    &39931       &8700\\
\hline
\end{tabular}
\label{example11}
\end{table}

\begin{table}[h]
\caption{Results on the BPBC Waveguide with a lossy core obtained by the MSEM and COMSOL}
\vskip0.2in
\centering
\begin{tabular}{ccccc}
\hline
$k_{i,z}^{N,h}$ &Eqn (\ref{eq:24})&Eqn (\ref{eq:25})&\textrm{COMSOL}& $k_{i,z}^{10,h}$\\
\hline
1&  1824046.06   & 1824045.86  & 1824046.12   & 1824046.03\\
 & -414641.90j   &-414641.81j  &-414641.91j   &-414641.88j\\

2&  1406129.71   & 1406128.43   & 1406129.82   & 1406129.58\\
 & -393456.61j   &-393451.08j  &-393457.14j   &-393457.07j\\

3&  1402755.24   & 1402751.41   & 1402756.63   & 1402756.43\\
 & -240078.01j   &-240074.79j  &-240078.56j   &-240078.43j\\

4&  1242473.47   & 1242473.60   & 1242473.42   & 1242473.41\\
 & -13859.87j    &-13859.94   &-13859.98j    &-13859.96j\\
\hline
Time(s)&5.979 & 7.135    &64.333     &28.230\\

DOF    &3132    &3132    &39931       &8700\\
\hline
\end{tabular}
\label{example12}
\end{table}

\subsection{A Patterned BPBC Waveguide for Future Application in Lithography}

\begin{figure}[h]
  \centering
{
   \includegraphics[width=0.45\columnwidth,draft=false]{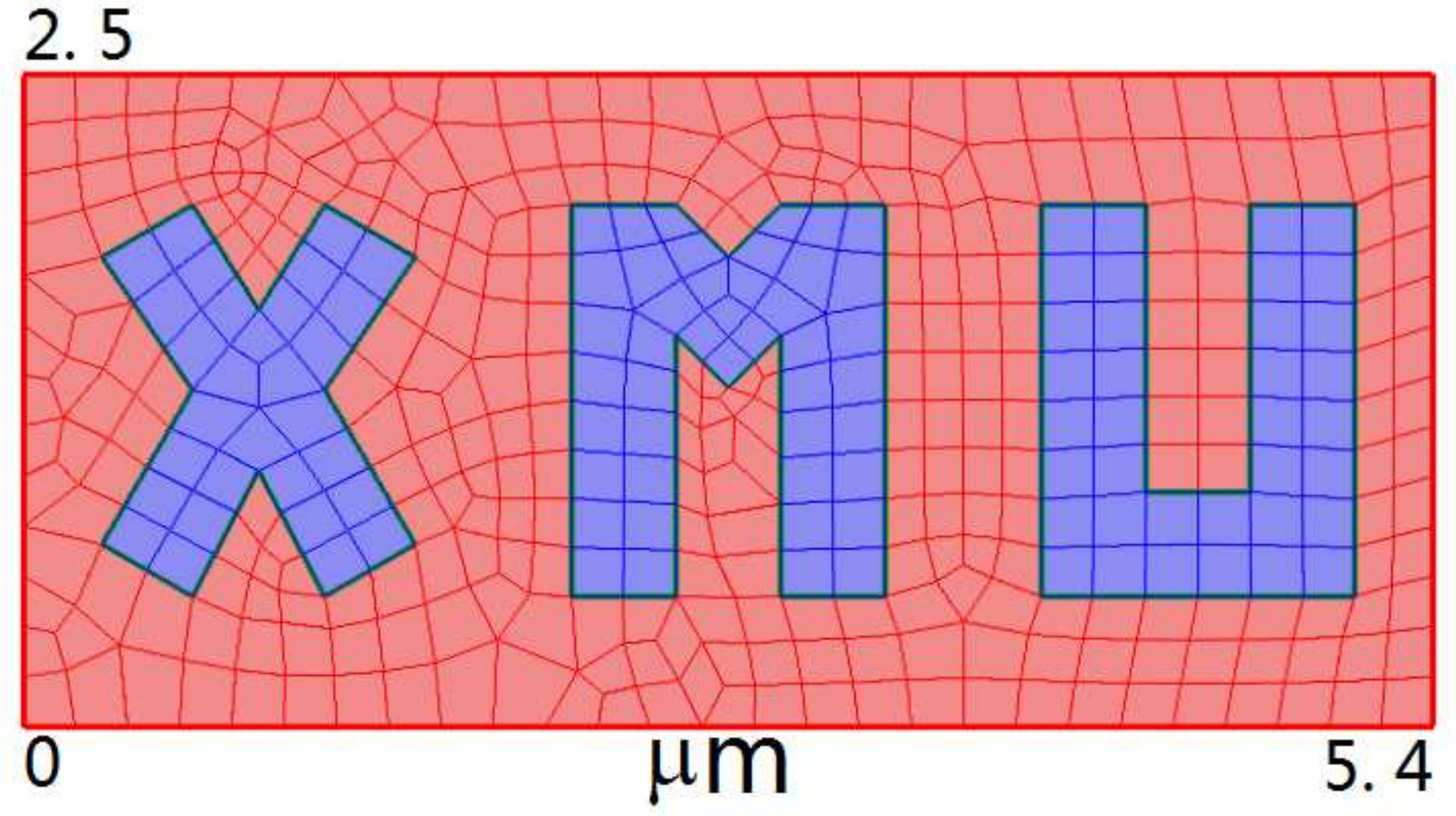}
   }
 \caption{Cross section for a patterned waveguide in lithography and its mesh for the MSEM.}
\label{sketch2}
\end{figure}

The lithography requires large scale computation to model light diffraction by optical masks.  One potential way to solve such a large scale problem is through the numerical mode matching method, where it is required to solve for the eigenvalues and eigenmodes of a large patterned optical waveguide problem with the BPBC.  One such example of a patterned waveguide cross section is shown in Fig. \ref{sketch2}. For the lithography, in order to obtain the electric and magnetic fields by using (\ref{eq:29}), we can solve a BPBC waveguide problem to achieve the propagation constant and corresponding modes. As shown in Fig. \ref{sketch2}, in this example the letters XMU with the size $0.4~\mu m$ are used as Bowtie holes and the dimensions of cross section are $5.4~\mu m \times 2.5~\mu m$.  The three letters X, M and U are set as the scatterers (Si, $\epsilon_{r}=12.0826$) in air. Here we assume that the light is perpendicularly incident, i.e., $(\theta,\phi)=(0,0)$. The operating wavelength is $1~\mu m$ and $k=k_{0}$.

\begin{figure}[h]
  \centering
{
   \includegraphics[width=0.6\columnwidth,draft=false]{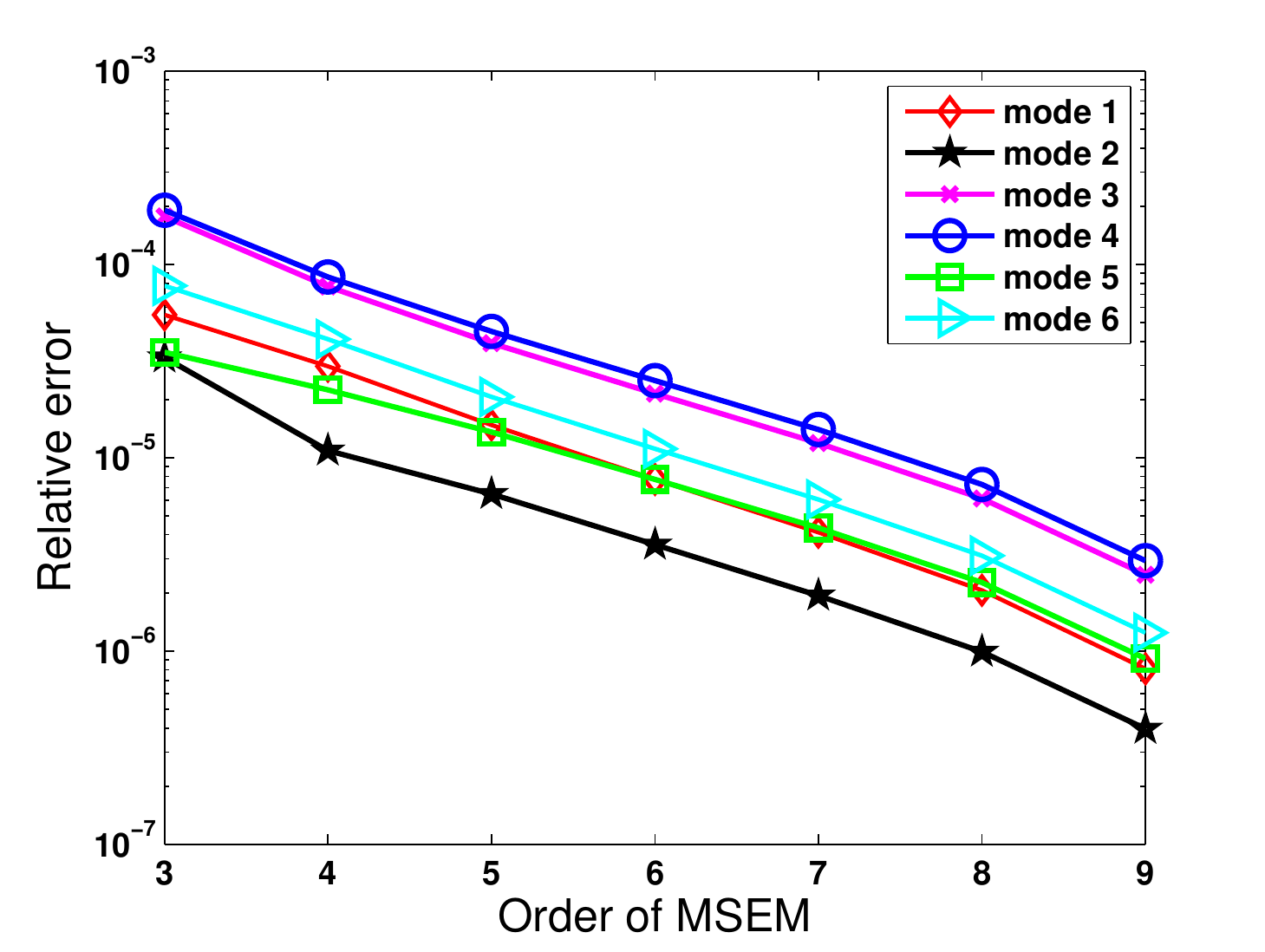}
   }
 \caption{Relative error $\textrm{log}(|k_{i,z}^{N,h}-k_{i,z}^{10,h}|/|k_{i,z}^{10,h}|)$ for Example 4.2.}
\label{error1}
\end{figure}

\begin{figure}[h]
\centering
\subfigure[]{
\label{value7}
\includegraphics[width=.3\textwidth]{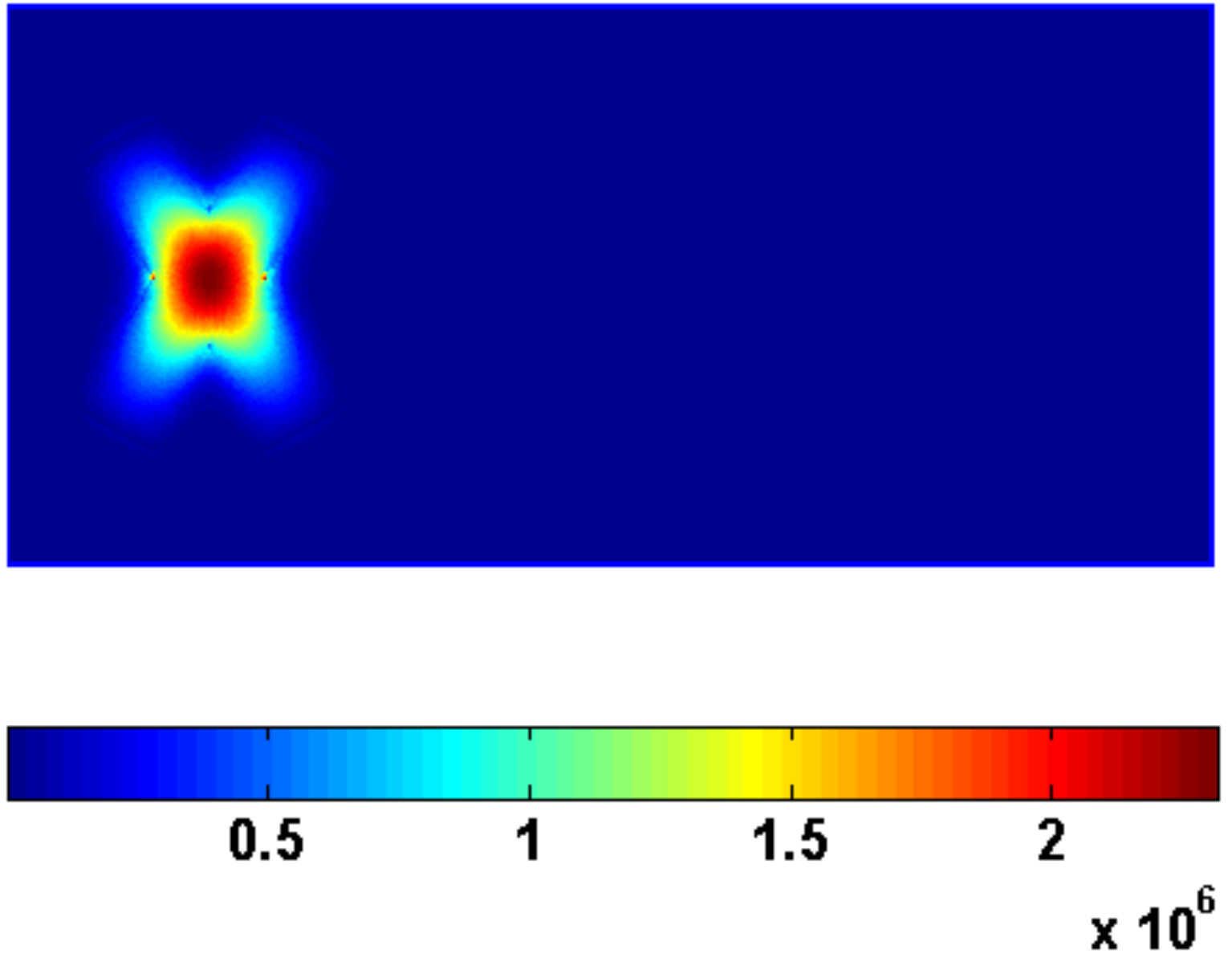}
}
\subfigure[]{
\label{value8}
\includegraphics[width=.3\textwidth]{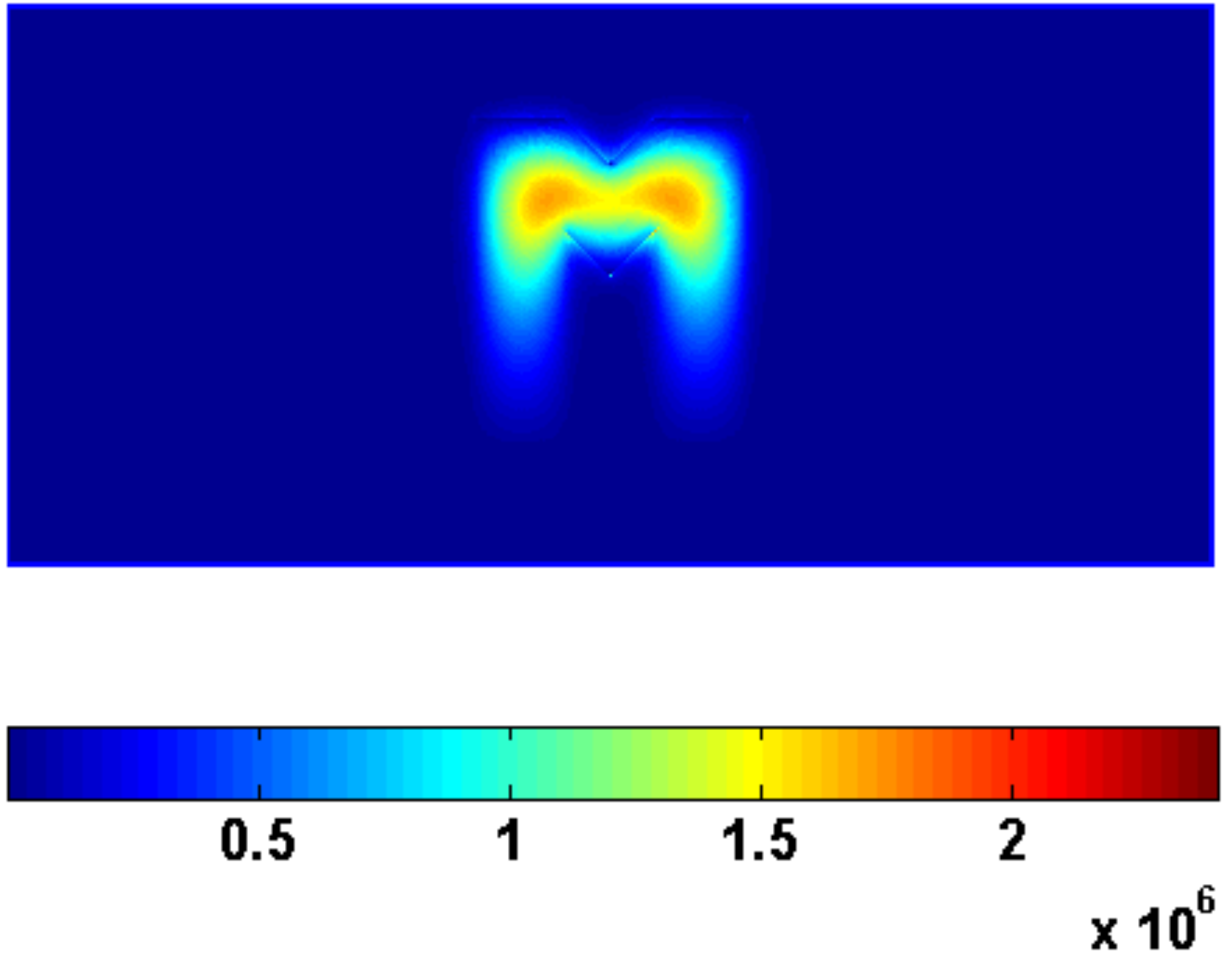}
}
\subfigure[]{
\label{value9}
\includegraphics[width=.3\textwidth]{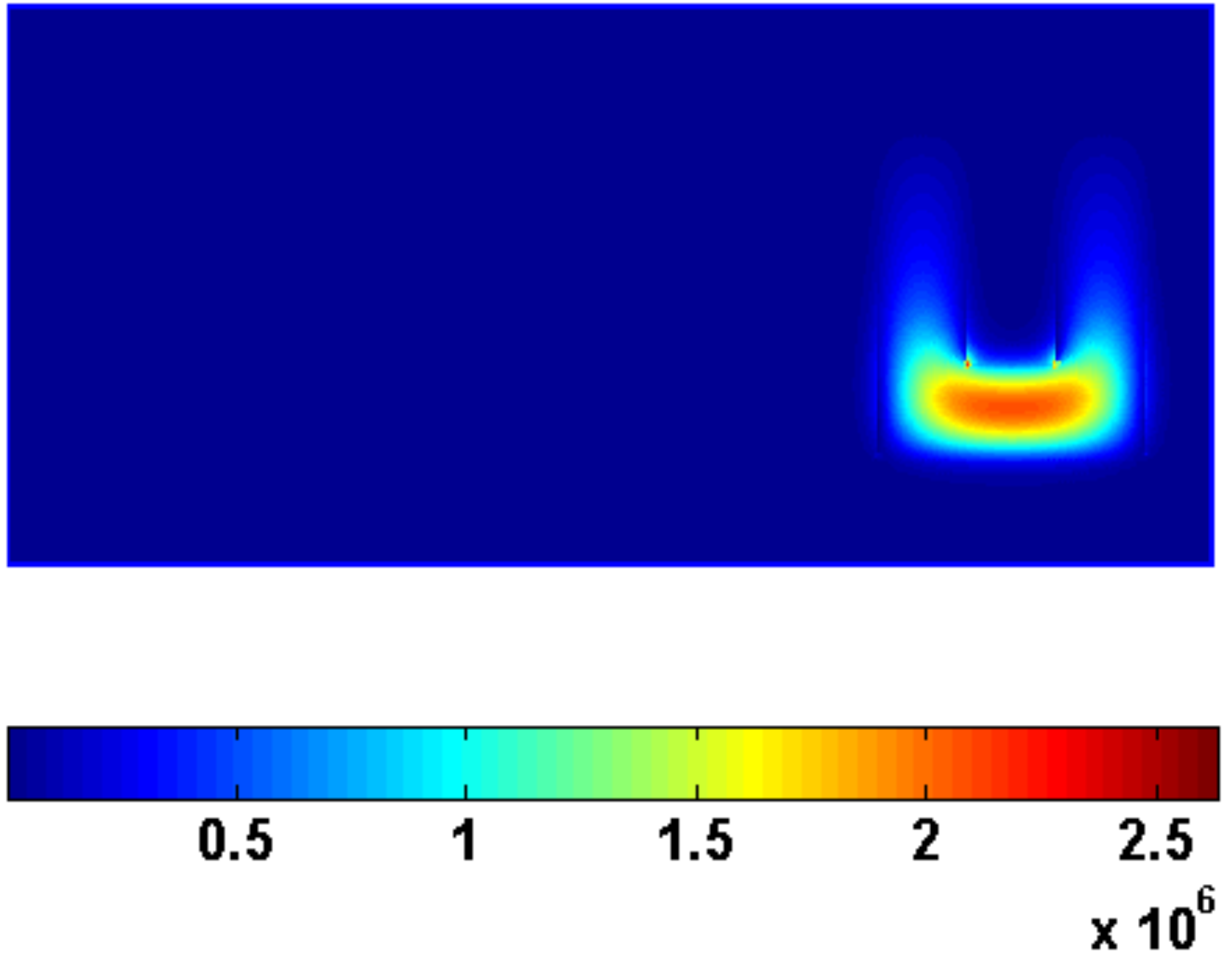}
}
\caption{Magnitude distributions of $\textbf{e}_{t}^{N,h}$. (a) The first mode. (b) The second mode. (c) The sixth mode.}
\label{field2}
\end{figure}

Here we apply the 3rd-order MSEM ($N=3$) and the 3rd-order FEM in COMSOL to solve (\ref{eq:24}) and (\ref{eq:25}) on the same mesh shown in Fig. \ref{sketch2}, respectively. The numerical eigenvalues are shown in Table \ref{example21}. We can see that from Table \ref{example21} the similar accurate solutions can be achieved by the MSEM and COMSOL, but COMSOL requires a little more DOFs and CPU time than MSEM even though our code is only an unoptimized research implementation. Meanwhile, it is easy to see that there are the same results for our two methods, which indicates that (\ref{eq:24}) and (\ref{eq:25}) are again equivalent. However, the high order modes may be required to the NMM method, in this case FEM will have a much lower accuracy than the MSEM with a large order ($N=10$). Table \ref{example22} indicates that in order to obtain similar accurate higher-order modes, COMSOL requires more 2.906 times DOFs and 2.346 times CPU time than the MSEM.
 The relative errors of the first four modes are plotted in Fig. \ref{error1} versus the order of MSEM basis functions, which show that the numerical results converge exponentially to the reference values with the order. The magnitude distributions of the field $\textbf{e}_{t}^{N,h}$ corresponding to the first, second, and sixth eigenvalues are also displayed in Fig. \ref{field2}. It clearly shows the three letters XMU.

\begin{table}[h]
\caption{Numerical results for the patterned waveguide in lithography}
\vskip0.2in
\centering
\begin{tabular}{ccccc}
\hline
$k_{i,z}^{N,h}$ &Eqn (\ref{eq:24})&Eqn (\ref{eq:25})&\textrm{COMSOL}& $k_{i,z}^{10,h}$\\
\hline
1&  21115512.108 &21115512.108 &21115231.113 &21114353.412\\

2&  21038407.237 &21038407.237 &21038315.975 &21038470.773\\

3&  21021691.678 &21021691.628 &21021213.174 &21017968.250\\

4&  20991922.939 &20991922.939 &20991523.142 &20987922.459\\

5&  20901959.925 &20901959.925 &20901952.096 &20901852.279\\

6&  20878861.638 &20878861.638 &20878511.349 &20877243.908\\
\hline
Time(s)&17.891 & 19.502    &22.333    &753.813\\

DOF    &10989    &10989    &11236       &122100\\
\hline
\end{tabular}
\label{example21}
\end{table}

\begin{table}[h]
\caption{Higher-order modes for the patterned waveguide in lithography}
\vskip0.2in
\centering
\begin{tabular}{ccccccc}
\hline
Solver &$k_{20,z}^{10,h}$&$k_{30,z}^{10,h}$ & $k_{40,z}^{10,h}$ &$k_{50,z}^{10,h}$& Time(s) & DOF\\
\hline
Eqn (\ref{eq:24}) &20239834.242& 19901700.521  & 19343724.159  & 18938248.569  & 759.813&  122100\\

Eqn (\ref{eq:25})&20239834.242&  19901700.521  & 19343724.159  & 18938248.569   & 801.297& 122100\\

\textrm{COMSOL} &20239834.621&  19901762.201   & 19343808.448  & 18938273.996   & 1782.667& 354775\\

\hline
\end{tabular}
\label{example22}
\end{table}

\subsection{BPBC Waveguide for Metasurface}

\begin{figure}[h]
  \centering
{
   \includegraphics[width=0.5\columnwidth,draft=false]{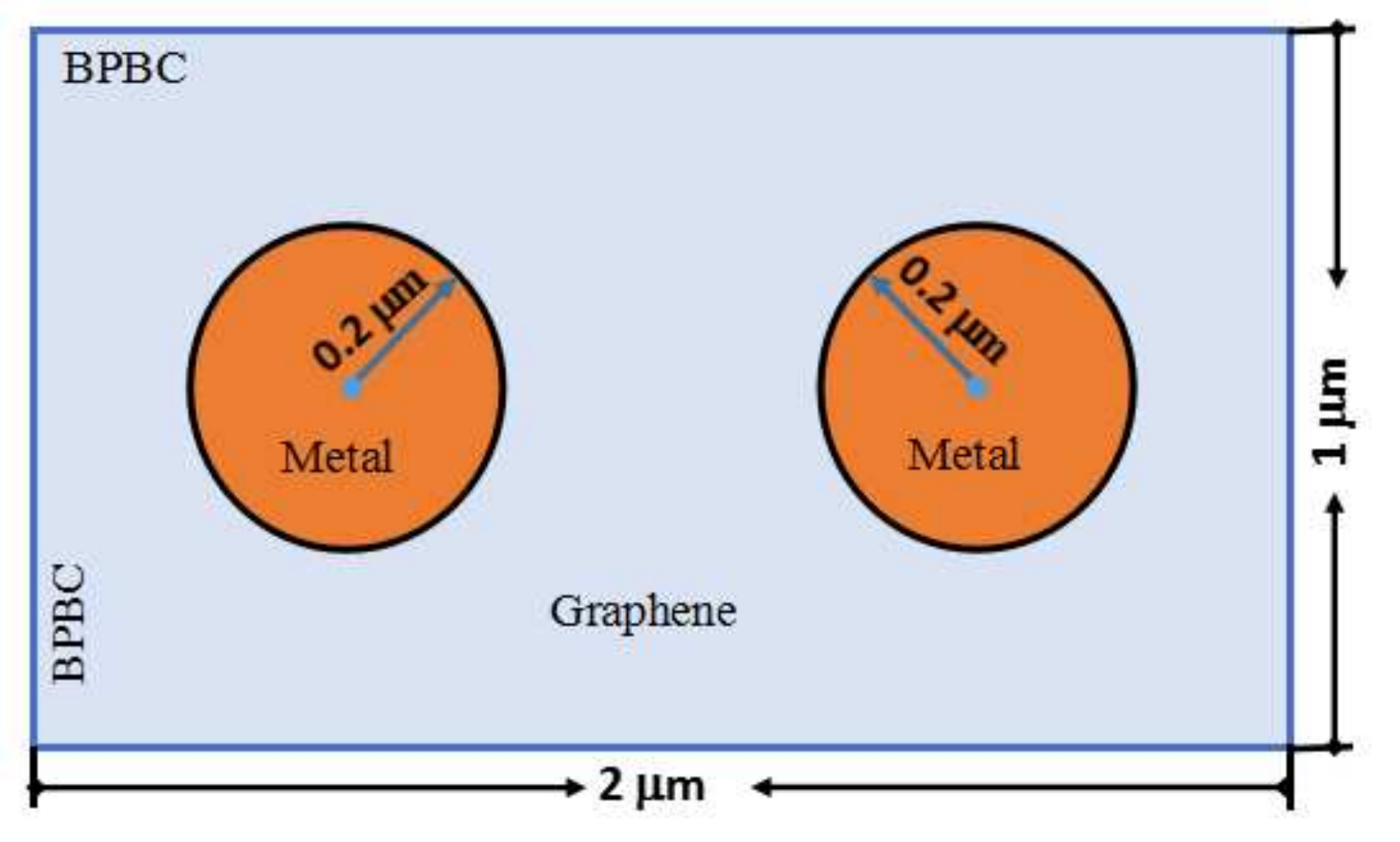}
   }
 \caption{A metasurface cross section with two metal objects.}
\label{sketch4}
\end{figure}

For metasurfaces, to calculate the refractive index and the reflectivity, again a potential approach by the NMM method \cite{QHLiu1990} is to first obtain the propagation constants and the waveguide modes by solving a waveguide problem with the Bloch periodic boundaries and the PEC boundaries shown in Fig. \ref{sketch4}, and then by using the form (\ref{eq:29}), the electric fields are obtained for calculating the refractive index and the reflectivity. Below we focus on calculating the propagation constants and the guided modes.

The rectangular BPBC waveguide has the dimensions $2~\mu m\times 1~\mu m$. The radius of the two circular metals is $0.2~\mu m$. The cross section of a metasurface is graphene whose the effective permittivity is $1-j\sigma/(\epsilon_{0}\omega t)$, where $\omega$ is angular frequency, $t=0.5~\textrm{nm}$ is the thickness and $\sigma$ is its conductivity with the parameters as given by \cite{Liu2016} and \cite{Francescato2013}. When the operating wavelength is taken as $6~\mu m$, the effective relative permittivity is $-37.4424-3.6711j$.
In this example, the elevation and azimuthal angles of the propagation direction are set as $(\theta,\phi)=(\pi/6,\pi/3)$, the 3rd-order MSEM ($N=3$) and the 3rd-order FEM are employed to solve (\ref{eq:24}) and (\ref{eq:25}) on a same mesh generated by COMSOL.

\begin{figure}[h]
  \centering
{
   \includegraphics[width=0.6\columnwidth,draft=false]{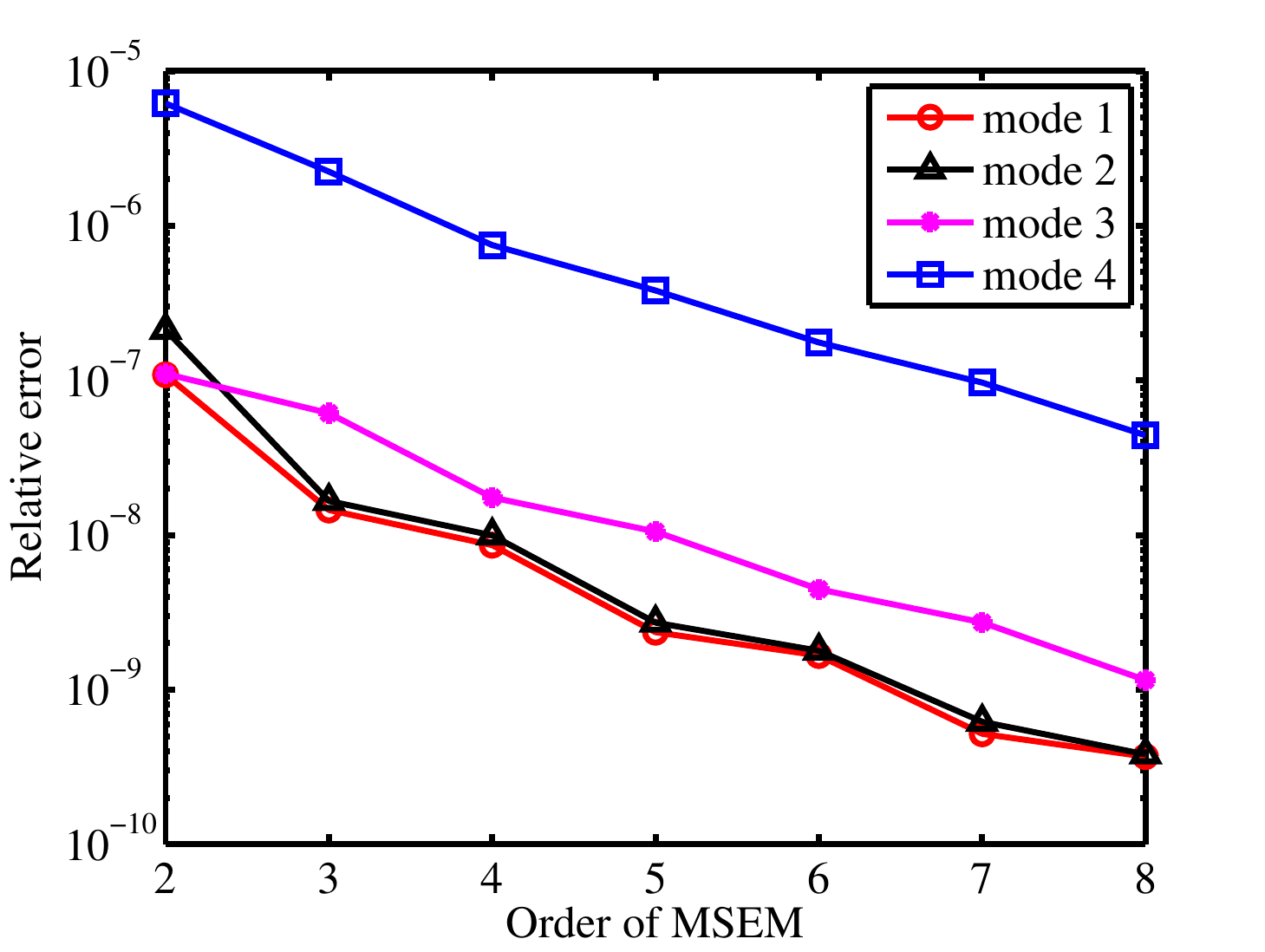}
   }
 \caption{Relative error $\textrm{log}(|k_{i,z}^{N,h}-k_{i,z}^{10,h}|/|k_{i,z}^{10,h}|)$ for Example 4.3.}
\label{error2}
\end{figure}

\begin{figure}[h]
  \centering
  \subfigure[]{
    \label{value9}
   \includegraphics[width=0.4\columnwidth,draft=false]{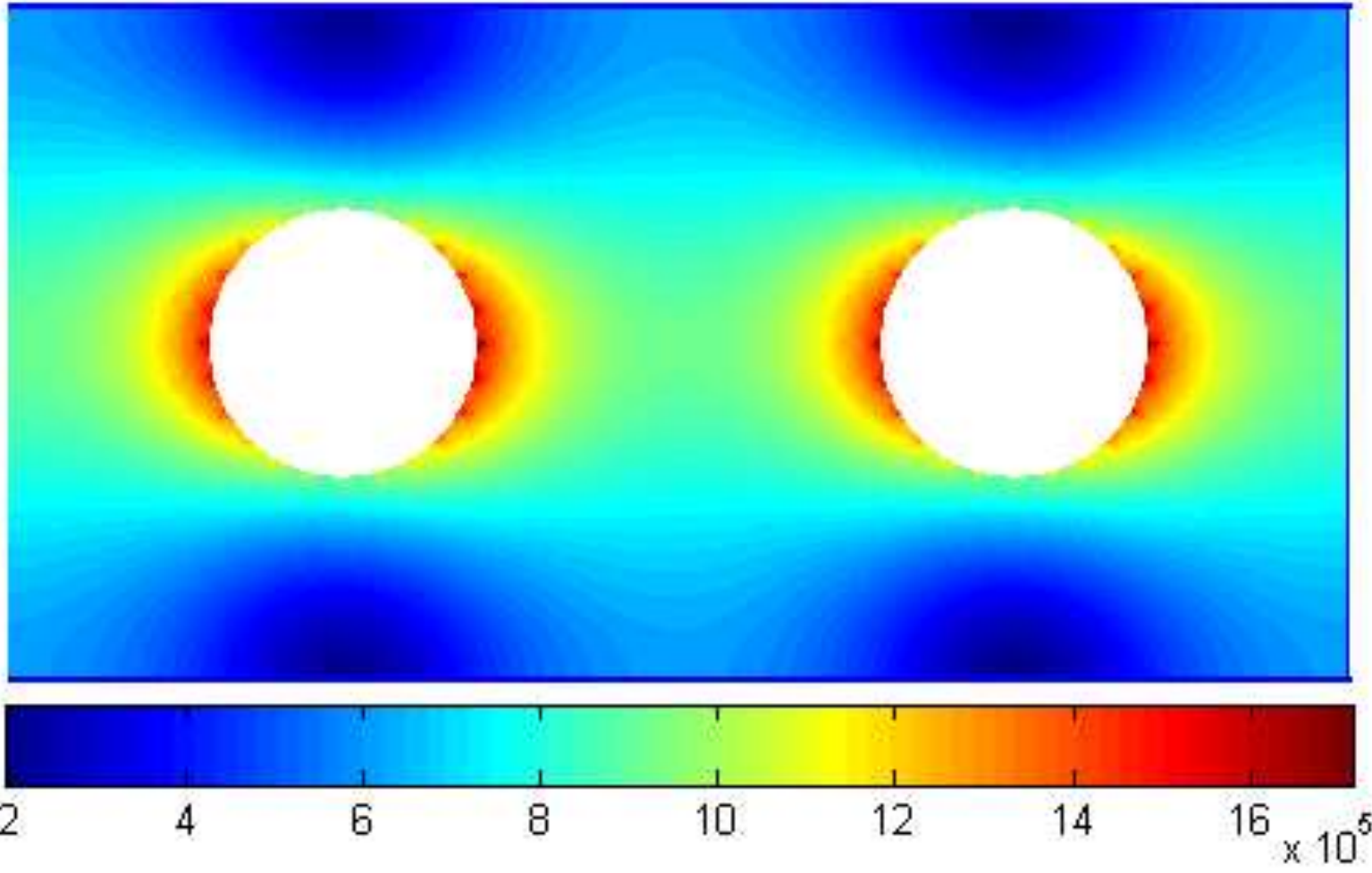}}
     \subfigure[]{
    \label{value10}
   \includegraphics[width=0.4\columnwidth,draft=false]{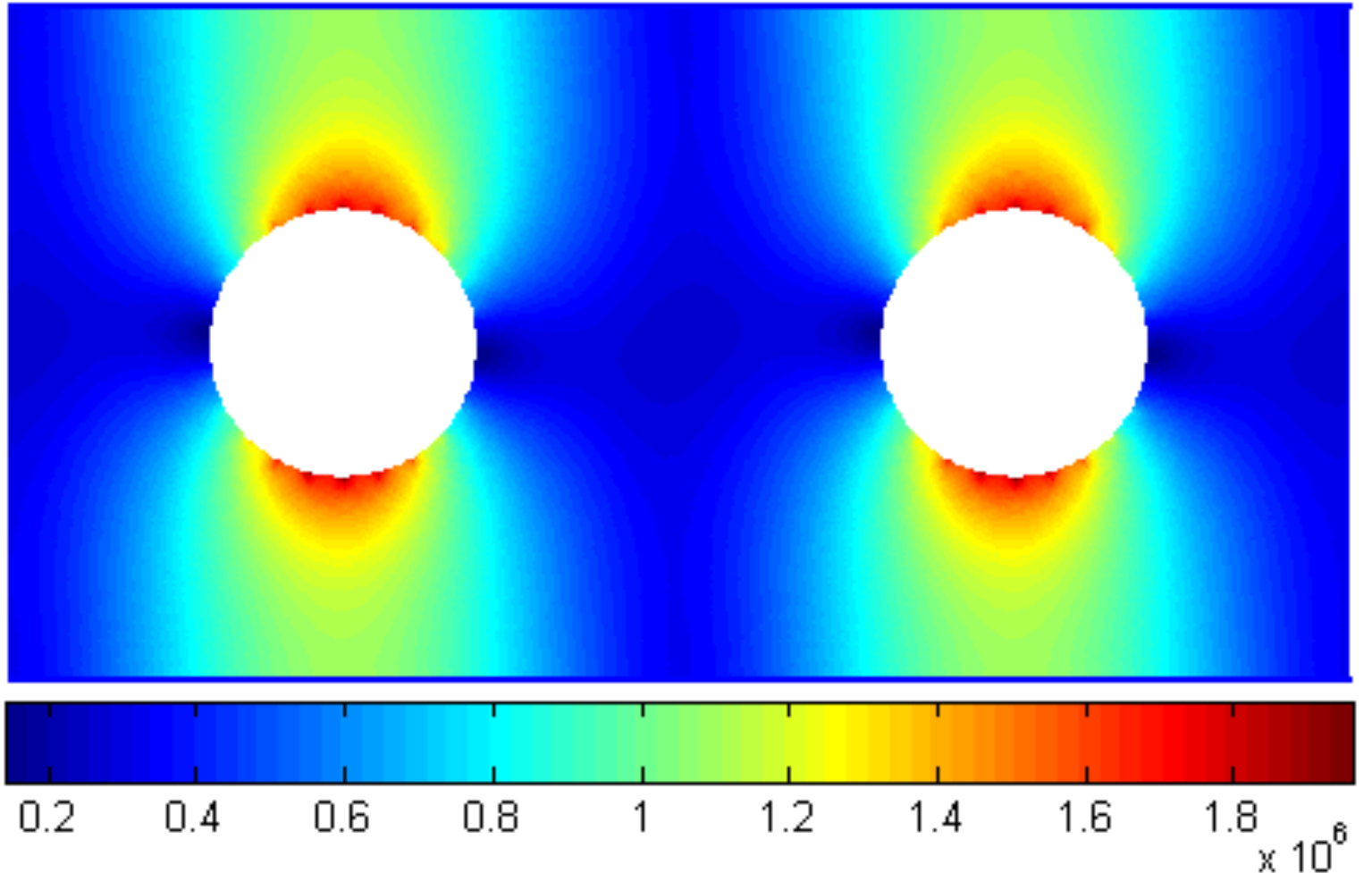}}
 \caption{Magnitude distributions of $\textbf{e}_{t}^{N,h}$ (a) The first mode. (b) The fourth mode.}
\label{field3}
\end{figure}

From Table \ref{example31}, we can see that there are two repeated propagation constants for this model, because there are two PEC boundaries. For the first three modes, COMSOL also obtains the nearly accurate solutions as the MSEM. Similarly, COMSOL spends more DOFs and CPU time than the MSEM. From Table \ref{example32}, to obtain the nearly accurate high order modes, COMSOL requires more 2.847 times DOFs and 8.659 times CPU time than the MSEM ($N=10$). The relative errors for the different order of the MSEM are depicted in Fig. \ref{error2}. It is shown that the first two modes converge exponentially to the reference value $k_{i,z}^{10,h}$ and the third mode follows. For the fourth modes, the rate of convergence is slower than the others, which is worse for COMSOL. The reason for this phenomenon is that the field has the saltuses around the PEC boundaries compared with the first mode (see, Fig. \ref{field3}).

\begin{table}[h]
\caption{Numerical results for BPBC waveguide corresponding to the metasurface}
\vskip0.2in
\centering
\begin{tabular}{ccccc}
\hline
$k_{i,z}^{N,h}$ &Eqn (\ref{eq:24})&Eqn (\ref{eq:25})&\textrm{COMSOL}& $k_{i,z}^{10,h}$\\
\hline
1&  313974.889  & 313974.885  & 313974.889   & 313974.889\\
 & -6419940.84j &-6419940.94j &-6419940.84j  &-6419940.84j           \\

2&  313974.889  & 313974.885  & 313974.889   & 313974.889\\
 & -6419940.84j &-6419940.95j &-6419940.84j  &-6419940.84j           \\

3&  313052.034  & 313051.866  & 313052.943   & 313051.847\\
 & -6438869.58j &-6438869.83j &-6438847.68j  &-6438870.23j           \\

4&  291604.892  & 291604.877  &291659.998 & 291604.223\\
 & -6912438.20j &-6912436.57j &-6911130.18j  &-6191245.21j          \\
\hline
Time(s)&8.125 & 9.166    &15.333    &60.503\\

DOF    &5902    &5902    &6371       &66718\\
\hline
\end{tabular}
\label{example31}
\end{table}

\begin{table}[h]
\caption{Higher-order modes for BPBC waveguide corresponding to the metasurface}
\vskip0.2in
\centering
\begin{tabular}{cccccc}
\hline
Solver &$k_{30,z}^{10,h}$ & $k_{40,z}^{10,h}$ &$k_{50,z}^{10,h}$& Time(s) & DOF\\
\hline
Eqn (\ref{eq:24}) &  169240.402  & 148109.876  & 136705.644   & 75.720&  66718\\
                  & -11910272.05j &-13609473.20j &-14744713.56  & &           \\

Eqn (\ref{eq:25})&  169240.423  & 148109.788  & 136705.648   & 92.112&  66718\\
                 & -11910276.64j &-13609500.38j &-14744820.32j  &  &           \\

\textrm{COMSOL} &  169146.673   & 148049.929  & 136696.382   & 655.667& 189959\\
                & -11917056.03j &-13617653.95j &-14746255.95j  & &           \\
\hline
\end{tabular}
\label{example32}
\end{table}

\section{Conclusion}

The mixed spectral element method (MSEM) is employed to solve the waveguide problem with the Bloch periodic boundary condition. Based on the Bloch periodic boundary condition (BPBC) and the periodic boundary condition (PEC), two equivalent mixed variational formulations are applied for the MSEM. Because the GLL curl-conforming vector basis functions are employed to discretize the variational formulations with the constraint of the Gauss' law, the MSEM is completely free of all the spurious modes and has the exponential convergence. A simple implementation method is used to deal with the BPBC and the PBC for the mixed variational formulations so that our MSEM schemes are not only easy to implement, but also can save computational costs. Three interesting examples are presented to verify that our schemes are more accurate and efficient than the finite element method.

\ack
 This research is partially supported by the National Natural Science Foundation of China under Grants 41390453 and 11501481, in part by the Key Scientific Project of Fujian Province in China under Grant 2015H0039 and in part by the Ph.D. Start-up Fund of the Natural Science Foundation of Guangdong Province, China, under Grant 2016A030310372.

\end{document}

%% file: latex_defs.tex
\def\^{\hat}
\def\~{\tilde}

\def\3h{{3\over 2}}

\def\eqn#1$${\eqno{{\rm #1}}$$}

\def\~{\tilde}

\def\^{\hat}

\def\bm#1{\mbox{{\boldmath${#1}$}}}

\def\XXint#1#2#3{{\setbox0=\hbox{$#1{#2#3}{\int}$}
     \vcenter{\hbox{$#2#3$}}\kern-.5\wd0}}
